\documentclass{aa}  

\usepackage{graphicx}
\usepackage{txfonts}
\usepackage{multirow}
\usepackage{hyperref}
\hypersetup{
    colorlinks=true,
    linkcolor=blue,
    urlcolor=blue,
    citecolor=blue
    }
\usepackage[dvipsnames]{xcolor}

\begin{document}

   \title{Metallicity dependence of the CO-to-H$_2$ and the [CI]-to-H$_2$ conversion factors in galaxies}
    \authorrunning{Bisbas et al.}
    \titlerunning{Metallicity dependence of conversion factors}

   \author{Thomas~G. Bisbas,
          \inst{1}
          Zhi-Yu Zhang,
          \inst{2,3}
          Maria-Christina Kyrmanidou,
          \inst{4}
          Gan Luo,
          \inst{5}
          Yinghe Zhao,
          \inst{6,7}
          Theodoros Topkaras,
          \inst{8}
          Xue-Jian Jiang,
          \inst{1}
          Donghui Quan,
          \inst{1}
          \and
          Di Li
          \inst{9,10}
          }

   \institute{Research Center for Astronomical Computing, Zhejiang Laboratory, Hangzhou, 311000, China\\
              \email{tbisbas@zhejianglab.com}
        \and
            School of Astronomy and Space Science, Nanjing University, Nanjing 210093, China\\
              \email{zzhang@nju.edu.cn}
        \and
            Key Laboratory of Modern Astronomy and Astrophysics (Nanjing University), Ministry of Education, Nanjing 210093, China
        \and
            Department of Physics, Aristotle University of Thessaloniki, Greece.
        \and
             Institut de Radioastronomie Millimetrique, 300 rue de la Piscine, 38400, Saint-Martin d'H\'eres, France
        \and
            Yunnan Observatories, Chinese Academy of Sciences, Kunming 650011, China
        \and
            Key Laboratory of Radio Astronomy and Technology (Chinese Academy of Sciences), A20 Datun Road, Chaoyang District, Beijing 100101, P. R. China
        \and 
            I. Physikalisches Institut, Universit\"at zu K\"oln, Z\"ulpicher Stra\ss e 77, D-50937 Köln, Germany
        \and
            Department of Astronomy, Tsinghua University, Haidian District, Beijing 100084, China
        \and
            National Astronomical Observatories, Chinese Academy of Sciences, Beijing 100101, China
             }

   \date{Received ---, accepted ---}

 
  \abstract
   {Understanding the molecular gas content in the interstellar medium (ISM) is crucial for studying star formation and galaxy evolution. The CO-to-H$_2$ ($X_{\rm CO}$) and the [C{\sc i}]-to-H$_2$ ($X_{\rm CI}$) conversion factors are widely used to estimate the molecular mass content in galaxies. However, these factors depend on many ISM environmental parameters such as metallicity, cosmic-ray ionization rate, and far-ultraviolet (FUV) radiation field, particularly in the low-metallicity ISM found in large galactocentric radii and in early-type galaxies.
   This work investigates the dependence of $X_{\rm CO}$ and $X_{\rm CI}$ on the ISM environmental parameters, with a focus on the low-metallicity $\alpha$-enhanced ISM ($\rm [C/O]<0$), to provide improved tracers of molecular gas in diverse conditions.
   We used the statistical algorithm {\sc PDFchem}, coupled with a database of photodissociation region (PDR) models generated with the {\sc 3d-pdr} astrochemical code. The models account for a wide range of metallicities, dust-to-gas mass ratios, FUV intensities, and cosmic-ray ionization rates. The conversion factors were computed by integrating the PDR properties over log-normal column density distributions ($A_{\rm V}$-PDFs) representing various cloud types.
   The $X_{\rm CO}$ factor increases significantly with decreasing metallicity, exceeding $\sim\!\!1000$ times the Galactic value at ${\rm [O/H] = -1.0}$ under $\alpha$-enhanced conditions, as opposed to $\sim\!\!300$ times under non-$\alpha$-enhanced conditions (${\rm [C/O]=0}$). In contrast, $X_{\rm CI}$ shows a more gradual variation with metallicity, making it a more reliable tracer of molecular gas in metal-poor environments under most conditions. The fraction of `CO-dark' molecular gas increases dramatically in low-metallicity regions, exceeding 90\% at ${\rm [O/H] = -1.0}$, particularly in diffuse clouds and environments with strong FUV radiation fields.
   The results highlight the limitations of CO as a molecular gas tracer in metal-poor ISM and demonstrate the potential of [C{\sc i}] (1-0) as a complementary tracer. The use of metallicity-dependent $X_{\rm CO}$ and $X_{\rm CI}$ factors, as provided by this study, is recommended for accurately estimating molecular gas masses in diverse environments. We recommend the use of the $\log_{10}X_{\rm CO}\simeq-2.41Z+41.3$ relation for the CO-to-H$_2$ conversion factor, and the $\log_{10}X_{\rm CI}\simeq-0.99Z+29.7$ relation for the [C{\sc i}]-to-H$_2$, where $Z=12+\log_{10}({\rm O/H})$.}


   \maketitle
%

\section{Introduction}
\label{sec:intro}

Determining the molecular gas content in galaxies is fundamental for understanding the evolution of the interstellar medium (ISM) and the star formation process across cosmological epochs. The H$_2$ molecule, however, is difficult to be directly observed in the cold ($T_{\rm gas}\sim10\,{\rm K}$) and dense ($n_{\rm H}\gtrsim10^3\,{\rm cm}^{-3}$) ISM state where stars form, its first two rotational transitions have upper-level energies of the order of $E/k_{\rm B}\simeq510\,{\rm K}$ and $1015\,{\rm K}$, respectively \citep{Dabrowski84}. Observations for the existence of the cold molecular gas rely, thus, on the emission captured by other tracers. In this regard, the molecule of $^{12}$CO (hereafter referred to as `CO') is perhaps the most commonly used, owing to its high abundance in molecular clouds \citep[$\sim\!\!10^{-4}$ at $Z=1\,{\rm Z_{\odot}}$][]{Sofia04}. Its $J=1-0$ transition has a low excitation energy barrier $\Delta E/k\sim5\,{\rm K}$ and low critical density of $\sim2.2\times10^3\,{\rm cm}^{-3}$ \citep{Bolatto13}, making it easily excited in cold environments \citep{Scoville16}. This emission line is then converted to an H$_2$ column density using the simple expression ${\rm N(H_2)}=X_{\rm CO}\cdot W_{\rm CO\,1-0}$ where $X_{\rm CO}$ is the conversion factor \citep[or `$X_{\rm CO}$-factor', see review by][]{Bolatto13} and $W_{\rm CO\,1-0}$ is the velocity integrated line emission of CO~(1-0). For typical Milky Way (MW) conditions, the value of $X_{\rm CO}=2\times10^{20}\,{\rm cm}^{-2}\,({\rm K}\,{\rm km}\,{\rm s}^{-1})^{-1}$ with a $\pm30$ per cent variation is usually adopted \citep{Strong96, Dame01, Bell07,Szucs16,Bisbas21,Skalidis24}.

The transition from atomic hydrogen to molecular hydrogen (frequently dubbed as `H{\sc i}-to-H$_2$ transition') is a vital first step in the process of star formation. When a cold ($T_{\rm gas}\lesssim30\,{\rm K}$) atomic cloud is sufficiently shielded from the interstellar ultraviolet radiation (UV) field, the H$_2$ molecule (mainly formed on dust grains) can no longer be photodissociated allowing the rapid increase of H$_2$ abundance, creating the conditions for the star-formation process to begin \citep[see review by][]{McKee07}. While extreme-ultraviolet (EUV) radiation ($h\nu\ge13.6\,{\rm eV}$) predominates in the vicinity of massive stars, far-ultraviolet (FUV) radiation ($6\lesssim h\nu<13.6\,{\rm  eV}$) governs the physical and chemical processes in photodissociation regions \citep[`PDRs', see reviews by][]{Hollenbach99,Tielens05,Wolfire22}. The H$_2$ formation rate coefficient and the dust absorption cross-section are both proportional to metallicity \citep{Sternberg14}. Therefore, as metallicity increases, the transition from H{\sc i}-to-H$_2$ shifts to lower column densities due to the increase in dust-to-gas mass ratio.

From the theory of PDRs, it is known that H$_2$ typically forms at lower visual extinctions than CO due to its self-shielding ability \citep{Tielens85}. There exists then a region which is H$_2$-rich but CO-poor \citep{Lada88}, leading to the so-called `CO-dark' molecular gas \citep{vDishoeck90,vDishoeck92}. The term `CO-dark' gas is to refer the H$_2$-rich gas not observable by the above line emission. Consequently, this has a direct impact on the accuracy in determining the H$_2$ gas mass through the CO~(1-0) millimeter line \citep[see e.g.][for recent developments]{Gong18,Gong20,Bigiel20,Luo20,Seifried20,Bisbas21,Dunne21,Dunne22,Hu22,Borchert22}.
Alternative tracers have been proposed in this regard. Amongst those, the [C{\sc i}] $^{3}P_1\rightarrow\,^{3}P_0$ at $609\,\mu$m (hereafter `[C{\sc i}]~(1-0)'; e.g. \citealt{Papadopoulos04}) and the [C{\sc ii}]~$158\mu$m \citep[e.g.][]{Zanella18,Zhao24} lines of the carbon cycle (C$^+$/C/CO) are of popular interest particularly due to the sensitivity of the Atacama Large Millimeter/submillimeter Array (ALMA) interferometry in these lines especially in observing high-redshift galaxies. 

The [C{\sc i}]~(1-0) line has been proposed as an alternative and accurate H$_2$ tracer more than two decades ago \citep{Papadopoulos04} as it is more optically thin than CO~(1-0) and has a low excitation temperature of $\Delta E/k\sim24\,{\rm K}$. The validity of using [C{\sc i}]~(1-0) in measuring the molecular gas mass has been confirmed by subsequent observational \citep{Lo14,Jiao17,Jiao19,Jiao21,Crocker19,Heintz20,Dunne21,Dunne22} and numerical works \citep{Offner14,Glover15,Glover16,Papadopoulos18,Gaches19b,Bisbas21}. Atomic carbon forms as long as the H{\sc i}-to-H$_2$ transition occurs \citep{Roellig07,Wolfire22} making the [C{\sc i}]~(1-0) line closely linked with the `CO-dark' gas. In the classical picture, the C-rich layer is always thought to be a thin layer \citep[e.g.][]{Tielens05}. However, recent theoretical models studying ISM environments with enhanced cosmic-ray energy densities \citep{Bialy15,Bisbas15,Bisbas17b} showed that the C-rich layer is a rather extended region covering a large density range consisted of cold H$_2$-rich gas. Three-dimensional PDR modeling \citep{Bisbas21} showed that the additional gas heating due to an elevated cosmic-ray ionization rate ($\zeta_{\rm CR}$) results in a [C{\sc i}]- and also [C{\sc ii}]-bright molecular gas. 


Both $^{12}$C and $^{16}$O are products of stellar nucleosynthesis albeit they are released in the ISM through different mechanisms \citep[see][for a review]{Romano22}. In particular, $^{16}$O is released through type-II supernova explosions from the massive $M\!\gtrsim\!8\,{\rm M}_{\odot}$, short-lived stars. On the other hand, $^{12}$C is mainly released through type-Ia supernovae of the low- and intermediate-mass stars ($1\!<\!M\!<\!8\,{\rm M}_{\odot}$) which are much longer-lived. Throughout the stellar evolution within a galaxy, the aforementioned differential release mechanisms cause a non-linear relationship between the abundances of C and O \citep{Trainor16,Cooke17,Berg16,Berg19,Maiolino19}. This was particularly studied in \citet{Berg19} who conducted chemical evolution models explaining that the observed trend of C/O versus O/H in metal-poor dwarf galaxies is a product of the detailed star formation history and supernova feedback. Environments that contain lower-than-expected C/O values  from the typical solar (linear) scaling are, therefore, characterized by the relative enrichment of oxygen, the most typical $\alpha$ element. Such environments are, thus, known as `$\alpha$-enhanced'.

It is not only the C/O abundance that scales differentially with O/H. The dust-to-gas (DTG) mass ratio is another important parameter across the cosmic ISM evolution \citep{Hunter24}. A lower DTG in metal poor ISM has been observed by various groups \citep[e.g.][]{Galametz11,Magdis11,Magdis12,Herrera12,Sandstrom13,RemyRuyer14,Shapley20,Galliano21,Popping22,Yasuda23,Konstantopoulou24} providing various scaling relations with O/H. Galaxy chemical evolution models show that dust grows via accretion of metals after a certain critical value of metallicity has been reached \citep{Asano13,Asano13b} while SNe~II increase the DTG during starburst events \citep{Zhukovska14}. These dust evolution approaches have been further explored in recent numerical models such as the {\sc DustySAGE} \citep{Triani20} and the `Feedback in Realistic Environments' (FIRE) galactic simulations of \citet{Choban24} providing a more complete picture of dust growth in various galaxy types. 

A more detailed impact of the aforementioned non-linear C/O and DTG ratios as a function of O/H on the abundances and line emission of the carbon cycle, was studied in \citet{Bisbas24}. Their three-dimensional PDR modeling showed that $\alpha$-enhanced metal-poor galaxies may appear to be bright in low-$J$ CO emission lines but dark in [C{\sc i}] lines \citep[in agreement with recent observations of {[}C{\sc i}{]}-dark galaxies, e.g.][]{Michiyama20,Michiyama21,Harrington21,Dunne22,Lelli23}, and possibly in the [C{\sc ii}]~$158\mu$m line. In addition, \citet{Bisbas24} find that in such $\alpha$-enhanced environments, low-$J$ CO lines appear to be the most stable ones for tracing molecular gas. The main conclusion of that work is that the C/O ratio cannot be neglected in astrochemical models as it constitutes an important ISM environmental parameter.

The impact of metallicity variations on the conversion factors has been already addressed and studied in various observations \citep[e.g.][]{Leroy11,Schruba12,Genzel12,Sandstrom13,Hunt15,Amorin16,Madden20} and numerical models \citep[e.g.][]{Accurso17,Bisbas21,Hu21,Ramambason24}. In general it is found that the $X_{\rm CO}$ conversion factor increases with decreasing metallicity with respect to the MW-average value, while it decreases towards higher metallicity regions, such as the Galactic Center \citep{Giveon02}.  
The kpc-scale hydrochemical models of \citet{Hu21} illustrate this trend (see their Figure~1). 
Recent observations of the low metallicity dwarf galaxy DDO~154 \citep{Komugi23} showed not only that the $X_{\rm CO}$ conversion factor is $\sim2-3$ orders of magnitude higher than the average in the MW, but an extended amount of CO-dark molecular gas may exist in that object.
Similar conclusions were addressed by \citet{Hunt23} in studying low-metallicity starburst galaxies, for which they find a difference between $5-10^3$ times of the Galactic $X_{\rm CO}$.
The recent radiative transfer models of \citet{Ramambason24} find that in metal-poor environments, the CO~(1-0) line becomes increasingly difficult to observe and thus not a good H$_2$-gas tracer. Instead, [C{\sc i}]~(1-0) and [C{\sc ii}] are good alternatives although the latter depends weaker on metallicity.

\begin{figure*}
    \centering
    \includegraphics[width=0.49\textwidth]{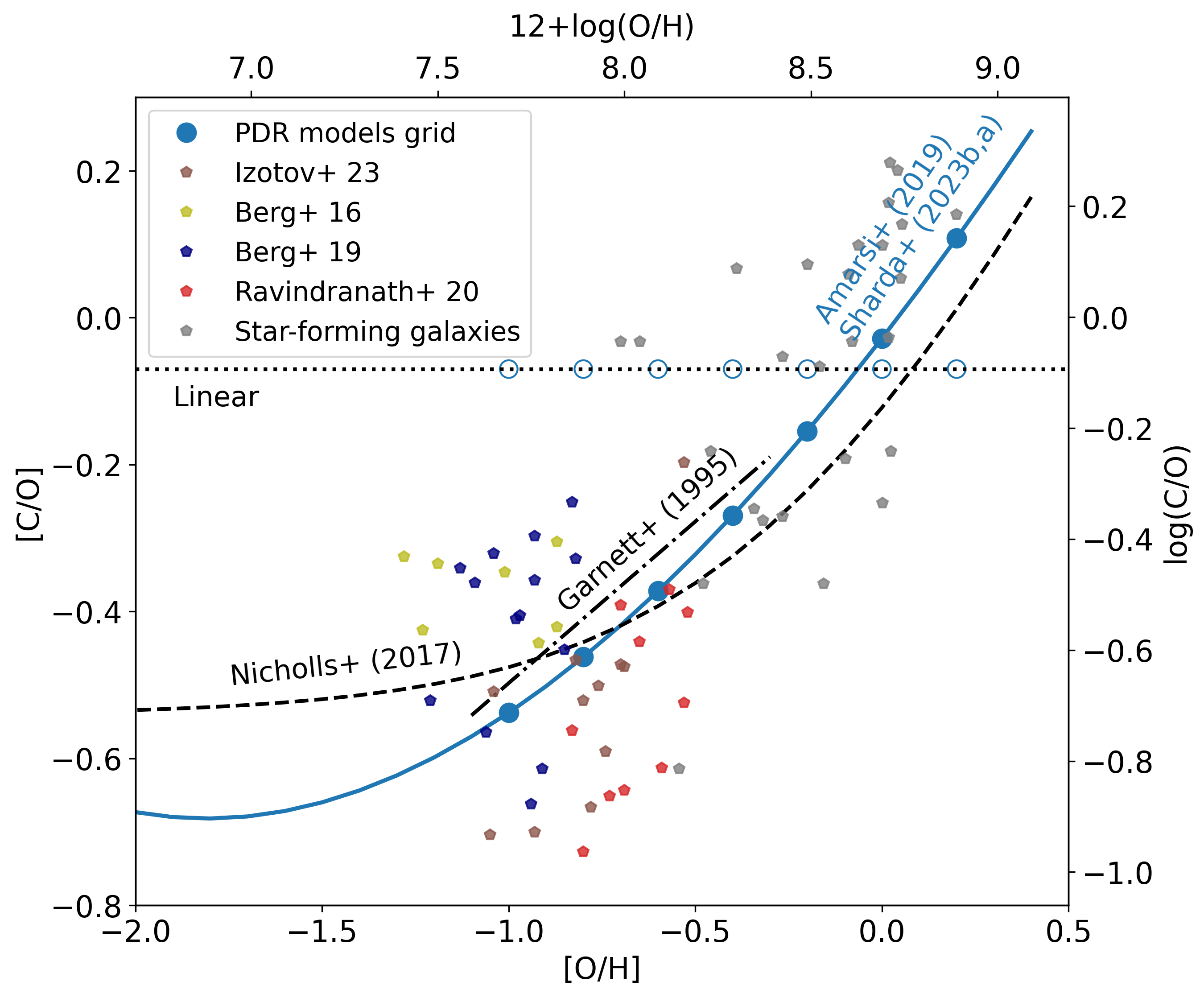}
    \includegraphics[width=0.49\textwidth]{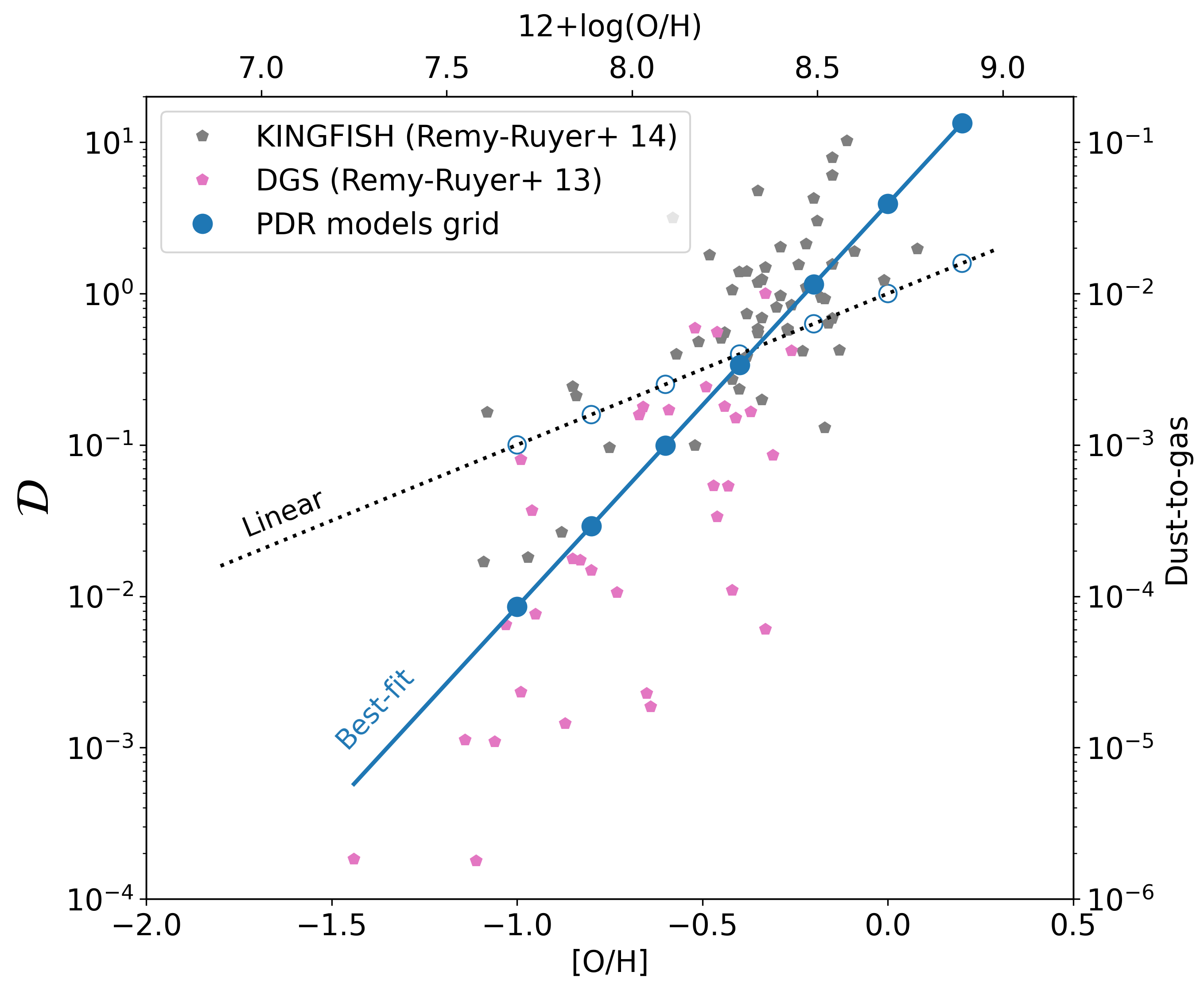}
    \caption{\textit{Left panel:} the [C/O]--[O/H] correlation based on \citet{Garnett95} (dot-dashed), \citet{Nicholls2017} (black dashed), and \citet{Amarsi19} observations, the latter using the best-fit relations of \citet{Sharda23,Sharda23cor} (blue solid). The dotted black line shows the case when C/H scales linearly with O/H. Filled blue circles show the [C/O]--[O/H] pairs for which we have performed the PDR modeling. The corresponding values for a linear relationship between C/H and O/H are shown in empty blue circles. The observational data of \citet{Berg16, Berg19, Ravindranath20, Izotov23} are shown for comparison, including star-forming galaxies \citep{Esteban02,Esteban09,Esteban14,Pilyugin05,Garcia-Rojas07,Lopez-Sanchez07}. The secondary y-axis shows the $\rm \log(C/O)$ abundance ratio. 
    \textit{Right panel:} The $\cal D$--[O/H] relation for the dust-to-gas ratio normalized to $10^{-2}$. Diamonds show observations from the DGS \citep{RemyRuyer13} and KINGFISH \citep{RemyRuyer14} surveys. The blue solid line shows the best-fit relation from these observations. The dotted black line shows the linear relationship between the dust-to-gas ratio and O/H. As with the left panel, the blue empty/filled circles show the inputted values to the PDR models. The secondary y-axis shows the non-normalized dust-to-gas mass ratio.}
    \label{fig:ics}
\end{figure*}

The scope of this work is to investigate how the CO-to-H$_2$ and the [C{\sc i}]-to-H$_2$ conversion factors depend on the ISM environmental parameters (the FUV intensity and the cosmic-ray ionization rate) in tracing the cold molecular gas in various metallicities, from metal-poor to metal-rich regions, with a particular focus on $\alpha$-enhanced ISM conditions. We do this by using the recently developed statistical tool {\sc PDFchem} \citep{Bisbas19,Bisbas23} to infer the PDR properties of column density distribution functions. This paper is organized as follows. In Section~\ref{sec:methods} we present the numerical approach we used. In Section~\ref{sec:results} we present the results of our analysis, followed by the relevant discussion in Section~\ref{sec:discussion}. We conclude in Section~\ref{sec:conclusions}.

\section{Numerical method}
\label{sec:methods}

\subsection{The {\sc PDFchem} algorithm}

For the purposes of this work, we use the newly developed and publicly available algorithm {\sc PDFchem}\footnote{\url{https://github.com/tbisbas/PDFchem}} \citep{Bisbas19,Bisbas23}. 
With {\sc PDFchem} we are able to estimate the average PDR properties, such as abundances and line emissions, for entire probability distribution functions of observed column densities or visual extinctions ($A_{\rm V,obs}$-PDF) from a pre-run PDR database. The PDR database used within {\sc PDFchem} is built with the publicly available code {\sc 3d-pdr}\footnote{\url{https://github.com/uclchem/3D-PDR}} \citep{Bisbas12}, which considers various thermochemical heating and cooling processes and terminates once thermal balance has been reached. 

During a {\sc PDFchem} run, the user imports an $A_{\rm V,obs}$-PDF function. The algorithm then performs weighted averages over that distribution to calculate the PDR properties from the database. These calculations are based on a set of pre-defined relations connecting the effective (or 3D/local) visual extinction, $A_{\rm V,eff}$, with the local H-nucleus number density, $n_{\rm H}$, and the observed visual extinction, $A_{\rm V,obs}$ \citep{Bisbas23}. In turn, these relations help to determine the depth up to which it is required to integrate for obtaining the column densities of species (hence the abundances) and also the line emission through additional radiative transfer. In particular, each PDR model uses a single and rather steep one-dimensional density slab (see Appendix~A of \citealt{Bisbas23}), which has been found to reproduce well, and at a minimal computational cost, the more complicated and computationally expensive three-dimensional PDR simulations. This density slab is embedded in various combinations of ISM environmental parameters. These include the strength of the FUV radiation field ($\chi/\chi_0=10^{-1}-10^3$ at a step of 0.1~dex and normalized to the spectral shape and flux of \citealt{Draine78}), the cosmic-ray ionization rate per H$_2$ molecule ($\zeta_{\rm CR}=10^{-17}-10^{-13}\,{\rm s}^{-1}$ at a step of 0.1~dex), as well as various combinations of O/H and C/O abundance ratios and dust-to-gas mass ratios described below. For all these PDR calculations, a chemical network of 33 species and 330 reactions has been used as a subset of the UMIST2012 database \citep{McElroy13}. The total number of PDR models used in this work is approximately $10^4$. 

\subsection{Abundances and dust-to-gas mass ratios}

The most crucial part of the PDR database used in this study is the choice of [O/H] and [C/O] ratios\footnote{The square bracket notation [A/B] represents the logarithmic ratio of the number of atoms of element A to element B, normalized to a reference value (here the Sun): 
\begin{eqnarray}
{\rm [A/B]}=\log_{10}\left(\frac{N_{\rm A}}{N_{\rm B}}\right)-\log_{10}\left(\frac{N_{\rm A}}{N_{\rm B}}\right)_{\odot},
\end{eqnarray}
\citep{Aller60,Pagel09}. The values $\rm [O/H]=0$ and $\rm [C/O]=0$ represent Solar abundances of O and C with $\rm 12+\log_{10}(O/H)=8.69$ and $\rm 12+\log_{10}(C/H)=8.43$, respectively \citep{Asplund2009}.} as well as the dust-to-gas ratios. The left panel of Figure~\ref{fig:ics} shows correlations between [C/O] and [O/H] found in the literature. The \citet{Garnett95} best-fit relation (black dot-dashed line) concerns {\it Hubble Space Telescope} observations of H{\sc ii} regions in dwarf galaxies and the Magellanic clouds. The \citet{Nicholls2017} (black dashed line) results from a collection of MW and nearby dwarf galaxies observations. The recent \citet{Sharda23,Sharda23cor} relation (solid blue line) is a best-fit in a sample of 187 stars \citep{Amarsi19} that exist in the MW thin and thick discs as well as the metal-poor halo. In this work, we follow the results of this relation. In addition, we also plot the observational data of metal-poor dwarf galaxies \citep{Berg16,Berg19}, low-redshift star-forming galaxies of low-metallicity \citep{Ravindranath20}, low-redshift Lyman-continuum leaking galaxies \citep{Izotov23}, and star-forming galaxies with abundances measured via recombination lines \citep{Esteban02,Esteban09,Esteban14,Pilyugin05,Garcia-Rojas07,Lopez-Sanchez07}. 

Our models consider the values of $\rm [O/H]=0.2,0,...,-1$ and the corresponding [C/O] values coming from the \citet{Sharda23,Sharda23cor} relation. 
In particular, the most representative object for each [O/H] value is as follows. The Galactic Center is found to have $Z\sim0.2\,{\rm Z}_{\odot}$ \citep{Giveon02} which best represents our $\rm [O/H]=0.2$ value. The chemical abundances of the Sun \citep{Asplund2009} and the depleted abundances in the solar neighborhood \citep{Sofia04} best represent the $\rm [O/H]=0$ and $-0.2$ values, respectively. The Large and Small Maggelanic clouds (LMC and SMC) \citep{Dufour84} are nearby low-metallicity extragalactic object, representative of our $\rm [O/H]=-0.4$ and $-0.6$ values. The outermost part of our Galaxy has also similar metallicities. The two lowest [O/H] values considered ($-0.8$ and $-1.0$) are to represent various metal-poor and extremely metal-poor dwarf galaxies \citep[e.g.][]{Shi15,Shi16}. The [C/O]-[O/H] pairs we mainly focus are marked with blue-filled circles. The dotted black line is to illustrate the correlation when a non-$\alpha$-enhanced (hereafter `linear') decrease of O/H and C/H is assumed in which the same decreasing factor is applied. With blue empty circles we plot the corresponding PDR models for that assumption.

The right panel of Fig.~\ref{fig:ics} shows the dust-to-gas ratio, $\cal D$, normalized to the solar neighbourhood value of $10^{-2}$ \citep{Sandstrom13}. The observations of the Dwarf Galaxies Survey \citep[`DGS'][]{RemyRuyer13} and the KINGFISH survey \citep[][]{RemyRuyer14} are shown while the blue solid line shows the corresponding best-fit relation. The black dotted line illustrates the linear assumption between $\cal D$ and [O/H]. As can be seen, the best-fit relation shows a super-linear correlation and it is given by the expression
\begin{eqnarray}
\label{eqn:dtg}
\log_{10}{\cal D} = \alpha\cdot{\rm [O/H]}+\beta
\end{eqnarray}
where $\alpha=2.66$ and $\beta=0.59$. Table~\ref{tab:ics} summarizes the input abundances and $\cal D$ values in the PDR models. In addition, we perform similar models but with the assumption of `linear' C/O and dust-to-gas ratio decrease (dotted black lines of panels in Fig.~\ref{fig:ics}). Details of these models are presented in Appendix~\ref{app:linear}.

\begin{table}[]
    \centering
    \begin{tabular}{c|c|c|c|l}
        [O/H] & $\cal D$ & C & O  & Comments\\ \hline
        0.2 & 13.376 & $5.49(-4)$ & $7.76(-4)$ & Galactic Center \\
       0 & 3.93 & $2.53(-4)$ & $4.90(-4)$ & Sun \\
        $-0.2$ & 1.15 & $1.19(-4)$ & $3.09(-4)$ & Solar neighb.\\
        $-0.4$ & 0.338 & $5.78(-5)$ & $1.95(-4)$ & LMC type\\
        $-0.6$ & 0.099 & $2.88(-5)$ & $1.23(-4)$ & SMC type\\
        $-0.8$ & 0.029 & $1.48(-5)$ & $7.77(-5)$ & Metal poor\\
        $-1.0$ & 0.0085 & $7.83(-6)$ & $4.90(-5)$ & Extr. metal poor\\
    \end{tabular}
    \caption{Initial conditions used in the grid of PDR models. The first column corresponds to the [O/H] value considered, the second to the $\cal D$ dust-to-gas ratio normalized to the value of $10^{-2}$ (see Eqn.~\ref{eqn:dtg}), the third to the carbon abundance and the fourth to the oxygen abundance \citep[calculated using the best-fit of][]{Sharda23,Sharda23cor}. In the fifth column we refer to the most representative objects in each case.}
    \label{tab:ics}
\end{table}

\begin{table}[]
    \centering
    \begin{tabular}{c|c|c|c}
        $m$ [mag] & $\sigma_{\rm PDF}$ & $\overline{A_{\rm V,obs}}$ [mag] & $\cal M$ [mag]\\ \hline
        \multirow{3}{*}{5} & 0.2 & 5.10 & 4.80\\
           & 0.4 & 5.42 & 4.26\\ 
           & 0.6 & 5.99 & 3.49\\\hline
        \multirow{3}{*}{10} & 0.2 & 10.20 & 9.60\\
           & 0.4 & 10.83 & 8.52\\
           & 0.6 & 11.97 & 6.98\\\hline
        \multirow{3}{*}{15} & 0.2 & 15.30 & 14.41\\
           & 0.4 & 16.25 & 12.78\\
           & 0.6 & 17.96 & 10.47\\
    \end{tabular}
    \caption{Summary of the properties of the inputted observed $A_{\rm V}$-PDFs in {\sc PDFchem}. The first column corresponds to the median value ($m$) of the $A_{\rm V}$-PDF, the second to the width of the $A_{\rm V}$-PDF ($\sigma_{\rm PDF}$), the third to the mean observed visual extinction ($\overline{A_{\rm V,obs}}$) and the fourth to the mode ($\cal M$) of the $A_{\rm V}$-PDF (see Equations 2-5 of \citealt{Bisbas19} for more details).}
    \label{tab:avpdfs}
\end{table}

\begin{figure}
    \centering
    \includegraphics[width=0.48\textwidth]{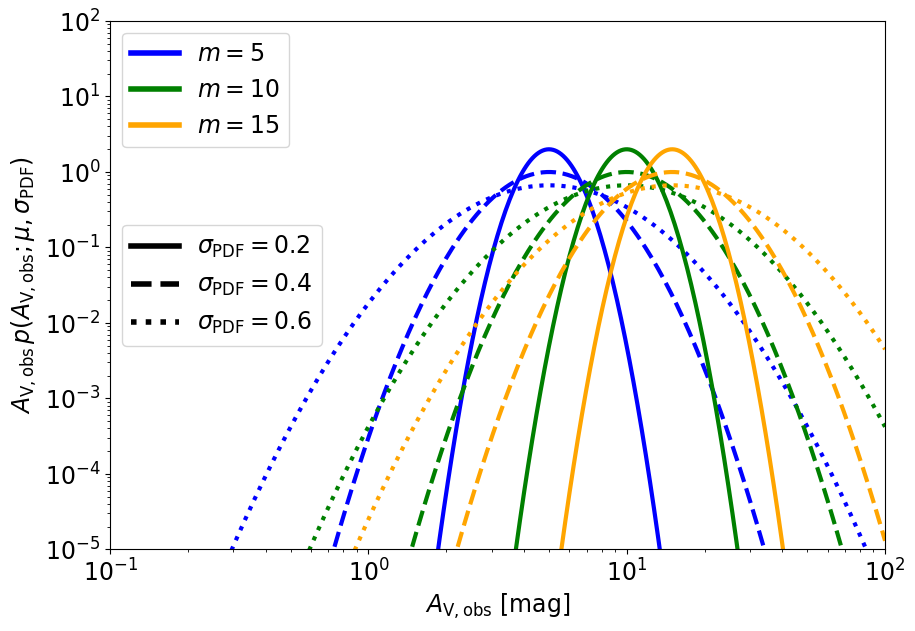}
    \caption{The inputted observed $A_{\rm V}$-PDFs used in {\sc PDFchem}. We used three different medians ($m=5$ blue line, $m=10$ green line, $m=15$ orange line) and three different widths ($\sigma_{\rm PDF}=0.2$ solid lines, $\sigma_{\rm PDF}=0.4$ dashed lines, $\sigma_{\rm PDF}=0.6$ dotted lines) corresponding to different values of the mean $\overline{A_{\rm V,obs}}$. Table~\ref{tab:avpdfs} summarizes the properties of these distributions.}
    \label{fig:avpdfs}
\end{figure}

\subsection{Column density distributions}

In regards to the inputted $A_{\rm V,obs}$-PDFs (hereafter simply referred to as `$A_{\rm V}$-PDFs') in {\sc PDFchem}, a collection of log-norm probability density functions is considered in which the median ($m$) and the width ($\sigma_{\rm PDF}$) are varied. In turn, this variation determines the values of the mean visual extinction ($\overline{A_{\rm V}}$) and the mode ($\cal M$), with the latter being the value of $A_{\rm V,obs}$ at which the distribution function peaks. We redirect the reader to \citet{Bisbas19} and their equations~(2)-(5) for their definition and connection to the probability function. Table~\ref{tab:avpdfs} presents a summary of the inputted $A_{\rm V}$-PDFs properties and Fig.~\ref{fig:avpdfs} their graphical representation. These distributions represent different types of ISM clouds that are dense enough to form stars under typical Milky-Way (MW) conditions. For example, \citet{Goodman09} report a mean $A_{\rm V,obs}\sim3\,{\rm mag}$ in clouds focusing on the Perseus star-forming region whose extent is of the order of $\sim20\,{\rm pc}$. \citet{Kainulainen09} report various mean $A_{\rm V}\sim1-3\,{\rm mag}$ of star-forming and non-star-forming clouds with sizes of $\sim4\,{\rm pc}$. Higher $A_{\rm V,obs}$ values ($\sim6\,{\rm mag}$) have been found by \citet{Froebrich10} in nearby giant molecular clouds, while a more extended study considering 195 nearby clouds has been performed by \citet{Abreu-Vicente15} covering cloud sizes of $\lesssim15\,{\rm pc}$ with mean $A_{\rm V,obs}\sim3-20\,{\rm mag}$. The latter range matches with the range considered in our work. Similar values have been found in the works of \citet{Schneider16, Spilker21, Ma22} and others \citep[see also][for further discussion]{Bisbas19, Bisbas23}.

For all metallicities explored, the above suite of $A_{\rm V}$-PDFs is treated as total column density distributions ($N_{\rm tot}$-PDFs) at solar metallicity. This assumption \citep[also taken previously in][]{Bisbas19,Bisbas21,Bisbas23,Bisbas24} implies then that the density distribution and the size of clouds are always kept the same across all metallicities, allowing for a better understanding on the trends of how conversion factors depend on the ISM environmental parameters. Under this assumption, the choice of $m=15$ was taken to ensure that considerable amounts of H$_2$-rich gas are always present in low-metallicities and under most of ISM environmental parameters explored. So for a given $A_{\rm V}$-PDF at any metallicity, we use the same $A_{\rm V,eff}-n_{\rm H}$ relationship in terms of density distribution and effective size of the cloud at solar metallicity, however integrating up to different depths (locations inside the cloud) depending on the metallicity.

\begin{figure*}
    \centering
    \includegraphics[width=0.98\textwidth]{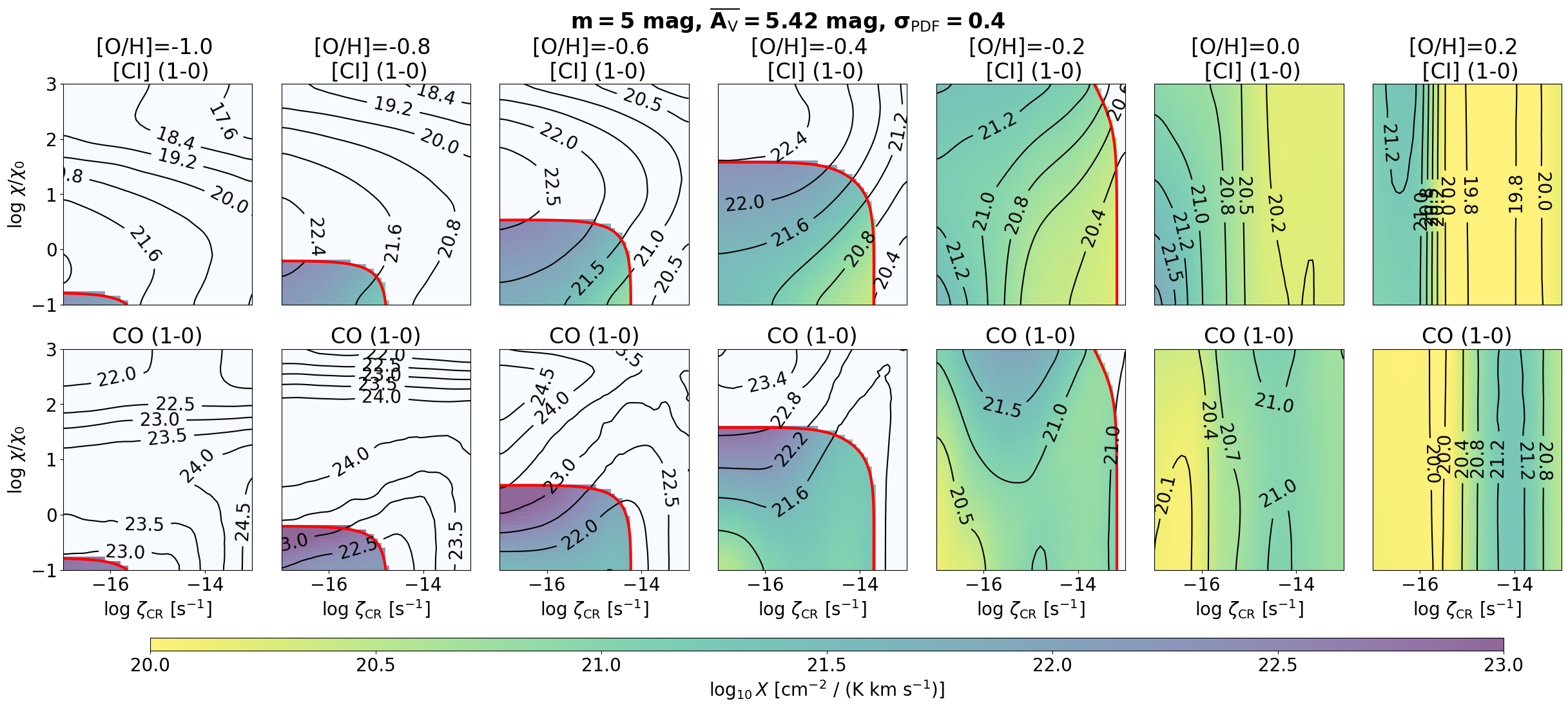}
    \caption{Logarithmic grid maps (in the common logarithmic space of $\log_{10}$) showing the $X$ conversion factors using the [C{\sc i}]~(1-0) (top) and the CO~(1-0) (bottom) defined as the ratio of the PDF-averaged H$_2$ column density to the PDF-averaged brightness temperature multiplied by the linewidth. These panels show results of the $A_{\rm V}$-PDF of $m=5\,{\rm mag}$ and $\sigma_{\rm PDF}=0.4$, corresponding to $\overline{A_{\rm V}}=5.42\,{\rm mag}$. Each column corresponds to different [O/H] ratios (left-to-right: $-1.0$ to $0.2$ at a step of $0.2$). In all panels, $x$-axis is the cosmic-ray ionization rate in the range of $\zeta_{\rm CR}=10^{-17} - 10^{-13}\,{\rm s}^{-1}$, and $y$-axis is the FUV intensity in the range of $\chi/\chi_0=10^{-1}-10^3$. The H{\sc i}-to-H$_2$ transition where the condition H{\sc i}~$=2$H$_2$ is satisfied, is marked in red solid line. For $\zeta_{\rm CR}-\chi/\chi_0$ pairs smaller than those corresponding to the red line the distribution is molecular (H$_2$-rich; colored partition), otherwise is atomic (H{\sc i}-rich; white partition). For $\rm [O/H]=0.0$ and $0.2$, the distribution is H$_2$-rich for the entire range of $\zeta_{\rm CR}$ and $\chi/\chi_0$ explored. The color bar shows the logarithm of the $X$-factor.}
    \label{fig:m5}
\end{figure*}

\subsection{Calculation of conversion factors}

We estimate the $X_{\rm C*}$ conversion factor of the coolant (where $\rm C*$ represents C or CO) using the following expression:
\begin{eqnarray}
  X_{\rm C*} = \frac{\langle \rm N(H_2)\rangle}{\langle W_{\rm C*}\rangle}\,\,\,[{\rm cm}^{-2}/({\rm K}\,{\rm km}\,{\rm s}^{-1})],
\end{eqnarray}
where $\langle \rm N(H_2)\rangle$ is the average H$_2$ column density of the $A_{\rm V,obs}$-PDF and $\langle W_{\rm C*}\rangle$ the average velocity integrated emission of the coolant. We estimate $\langle \rm N(H_2)\rangle$ using the following expression:
\begin{eqnarray}
    \langle {\rm N(H_2)}\rangle = \frac{\sum_i {\rm N(H_2)}_i \times {\rm (PDF \cdot \Delta A_{\rm V})}_i}{\sum_i \,{\rm (PDF \cdot \Delta A_{\rm V})}_i},
\end{eqnarray}
where $\rm N(H_2)_i$ is the H$_2$ column density of the $i$-th computational element of the $A_{\rm V,eff}-n_{\rm H}$ relation \citep[see][]{Bisbas19,Bisbas23}.

For each PDR model and depth point, we calculate the radiation temperature ($\rm T_{\rm C*}$) of the coolant by solving the radiative transfer equation as explained in \citet{Bisbas17b,Bisbas23}. For the calculation of the optical depth and consequently the radiation temperature, we use the level populations of the coolants resulting from each {\sc 3d-pdr} model under LTE conditions and using the LVG escape probability approximation \citep[see][for full details]{Bisbas12}.

The average velocity integrated emission, $\langle W_{\rm C*}\rangle$, is estimated by calculating $\rm T_{C*}$, multiplied by the linewidth, $\Delta V$:
\begin{eqnarray}
    \langle W_{\rm C*}\rangle = \frac{\sum_i {\rm T}_{{\rm C*},i}\,\Delta V_{i} \times {\rm (PDF \cdot \Delta A_{\rm V})}_i}{\sum_i \,{\rm (PDF \cdot \Delta A_{\rm V})}_i},
\end{eqnarray}
where 
\begin{eqnarray}
    \Delta V = 2\sqrt{2\ln2}\sqrt{\frac{k_{\rm B}T_{\rm gas}}{m_{\rm mol}}+v_{\rm Turb}^2}\,\,\,{\rm [km/s].}
\end{eqnarray}

In all models presented here, we assume a macroturbulent velocity dispersion of $v_{\rm Turb}=1.5\,{\rm km/s}$. For instance, for the CO molecule and for a gas with $T_{\rm gas}=10\,{\rm K}$, the above expression gives a $\Delta V\sim3.5\,{\rm km}\,{\rm s}^{-1}$, in agreement with observations \citep[e.g.][]{Heyer15,Rice16}.

\section{Results}
\label{sec:results}

\subsection{Behaviour of $X$-factors on the ISM environmental parameters}
\label{ssec:detailed}

\begin{figure*}
    \centering
    \includegraphics[width=0.98\textwidth]{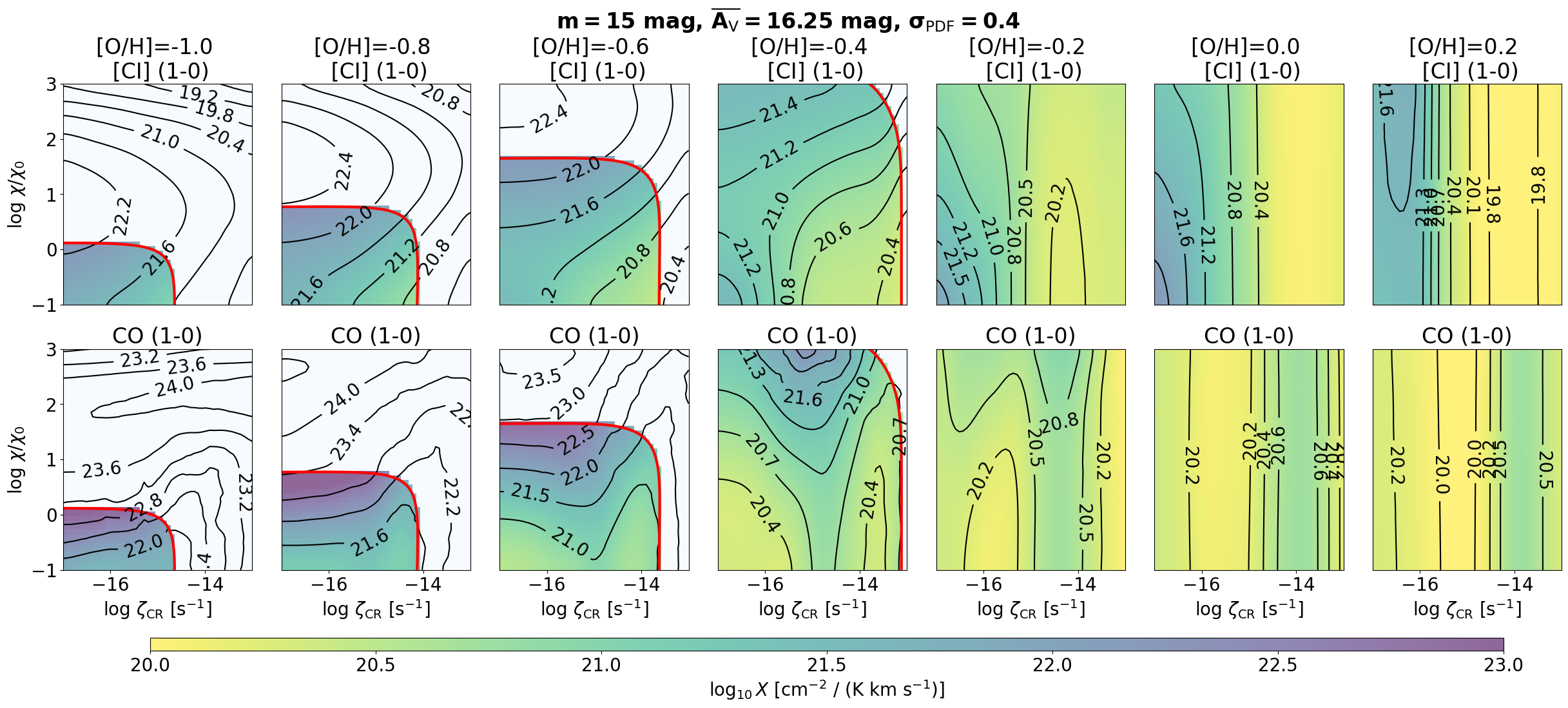}
    \caption{As in Fig.~\ref{fig:m5} but for an $A_{\rm V}$-PDF distribution of $m=15\,{\rm mag}$ and $\sigma_{\rm PDF}=0.4$ corresponding to an average $A_{\rm V}$ of $\overline{A_{\rm V}}=16.25\,{\rm mag}$. For $\rm [O/H]=-0.2$ to $0.2$, the distribution remains H$_2$-rich for the entire range of $\zeta_{\rm CR}$ and $\chi/\chi_0$ explored.}
    \label{fig:m15}
\end{figure*}

Figures~\ref{fig:m5} and \ref{fig:m15} show logarithmic grid maps\footnote{In Figures~\ref{fig:m5}, \ref{fig:m15}, and \ref{fig:ratios}, we apply a two-dimensional \citet{Savitzky64} filter to reduce the numerical noise, using the algorithm available at \url{https://github.com/espdev/sgolay2}.} of the $X$-factors for two representative $A_{\rm V}$-PDFs, namely with $m=5$ and $15\,{\rm mag}$ and with $\sigma_{\rm PDF}=0.4$. Each row shows the $\log_{10}X$ factor for the two tracers ([C{\sc i}]~(1-0) and CO~(1-0)). Each column corresponds to different [O/H] with the corresponding [C/O] and $\cal D$ presented in Table~\ref{tab:ics} (see also Fig.~\ref{fig:ics}). The red solid line corresponds to the H{\sc i}-to-H$_2$ transition, meaning that for lower $\zeta_{\rm CR}-\chi/\chi_0$ pairs than that of the red line (colored partition), the $A_{\rm V}$-PDF is H$_2$-rich. In general, this transition decreases in the $\zeta_{\rm CR}-\chi/\chi_0$ space as [O/H] decrease, so the distribution remains molecular for weak values of FUV intensities and cosmic-ray ionization rates. For $\rm [O/H]\ge0$, the distributions remain molecular for most/all of the $\zeta_{\rm CR}-\chi/\chi_0$ range. The particular set of $m=5$ distributions are mainly atomic for $\rm [O/H]=-1.0$. As can be expected, we find that log-normal $A_{\rm V}$-PDFs with $m<5\,{\rm mag}$ hardly contain any molecular gas in low metallicity ISM conditions. 

\begin{figure}
    \centering
    \includegraphics[width=0.9\linewidth]{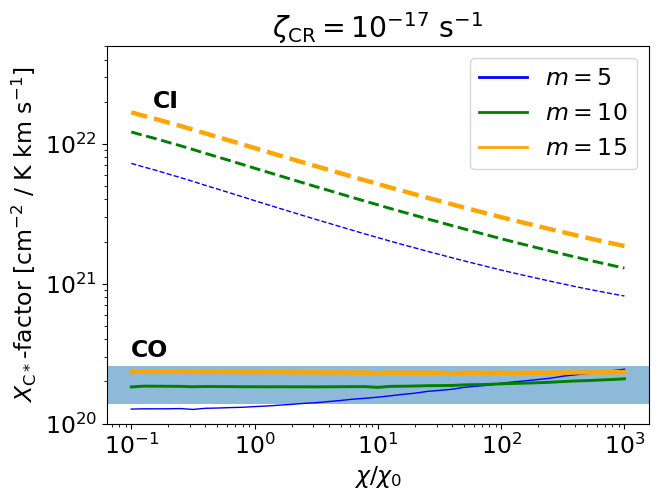}
    \includegraphics[width=0.9\linewidth]{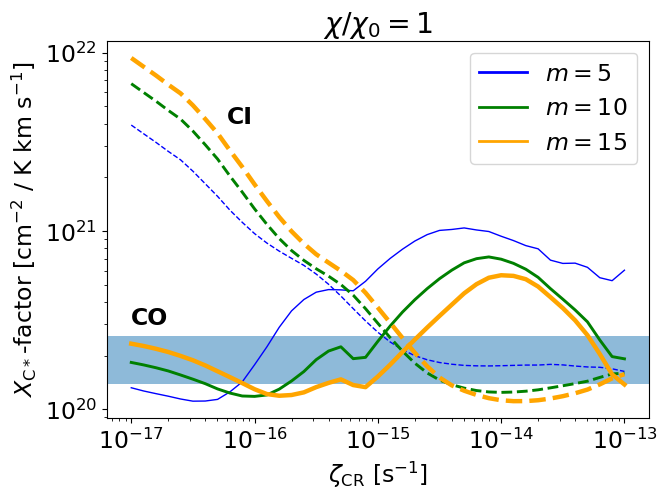}
    \caption{Conversion factors at $\rm [O/H]=0$ for a fixed $\zeta_{\rm CR}=10^{-17}\,{\rm s}^{-1}$ (top panel) and for a fixed $\chi/\chi_0=1$ (bottom panel). The $A_{\rm V}$-PDFs have all a width of $\sigma_{\rm PDF}=0.4$. Blue lines correspond to $m=5$, green lines to $m=10$ and orange lines to $m=15\,{\rm mag}$. Solid lines correspond to the $X_{\rm CO}$ and dashed to the $X_{\rm CI}$. The blue-shaded stripe represents the average MW value.}
    \label{fig:fixed}
\end{figure}

In terms of the dependency of each $X$-factor on $\zeta_{\rm CR}$ and $\chi/\chi_0$ and the corresponding trends, we can get an insight from both Figs.~\ref{fig:m5} and \ref{fig:m15}. The $X_{\rm CI}$-factor illustrated in the top row of both figures shows almost no dependency on the intensity of the FUV radiation field for $\rm [O/H]\ge0$ but rather on the cosmic-ray ionization rate, since the high $\cal D$ value shields the distribution from the interaction with FUV photons. We find that $X_{\rm CI}$ decreases with increasing $\zeta_{\rm CR}$ due to the effect of cosmic-ray induced destruction of CO, leading to an increase in C abundance \citep{Bisbas15,Bisbas17} and consequently to the [C{\sc i}](1-0) emission \citep{Bisbas21}. As both [C/O] and $\cal D$ decrease with [O/H] as illustrated in Fig.~\ref{fig:ics}, the FUV radiation is able to penetrate further reaching higher column densities and consequently higher H-nucleus number densities. This makes the $X_{\rm CI}$-factor to become more responsive to the change of $\chi/\chi_0$ value. For solar neighborhood metallicities and above ($\rm [O/H]\gtrsim-0.2$), and for a low and fixed $\zeta_{\rm CR}$, the $X_{\rm CI}$ decreases with increasing intensity of the FUV radiation field, as we illustrate in a later figure (Fig.~\ref{fig:fixed}). This occurs due to the faster increase of [C{\sc i}] emission with respect to the slower decrease of N(H$_2$).

The bottom row of Figs.~\ref{fig:m5} and \ref{fig:m15} shows the dependency of $X_{\rm CO}$ on the $\zeta_{\rm CR}$ and $\chi/\chi_0$ for the two representative $A_{\rm V}$-PDFs. As with the $X_{\rm CI}$, the CO conversion factor depends more strongly on $\zeta_{\rm CR}$ than on the FUV intensity for $\rm [O/H]\ge0$ and even for $\rm [O/H]=-0.2$ in high-density clouds. For more diffuse clouds than explored here, it is expected that at those metallicities the $X_{\rm CO}$ will inevitably have a dependency on the $\chi/\chi_0$ value \citep[see e.g.][]{Bisbas23}. Contrary to the trends observed in $X_{\rm CI}$, the $X_{\rm CO}$-factor shows a different behavior as $\zeta_{\rm CR}$ increases for a fixed $\chi/\chi_0$; $X_{\rm CO}$ fluctuates irregularly as the cosmic-ray ionization rate increases. Note that the discrepancy in $X_{\rm CO}$ values, especially in the $m=15\,{\rm mag}$ distribution (Fig.~\ref{fig:m15}) for $\rm [O/H]\gtrsim-0.2$, is very small. This implies that for such high column density clouds, the $X_{\rm CO}$-factor depends weakly on the ISM environmental parameters. For lower metallicities, the aforementioned conversion factor depends mainly on the intensity of the FUV radiation even for the higher column density distribution. It is expected that cosmic-rays will have an impact in low metallicity environments only for the very high density condensations where FUV photons cannot penetrate.

\begin{figure*}
    \centering
    \includegraphics[width=0.49\textwidth]{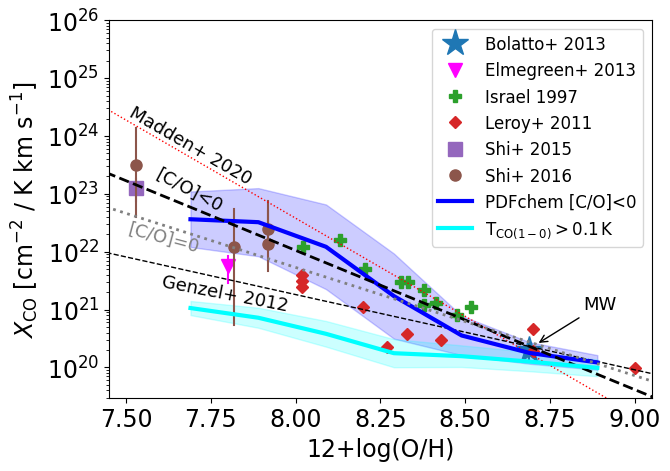}
    \includegraphics[width=0.49\textwidth]{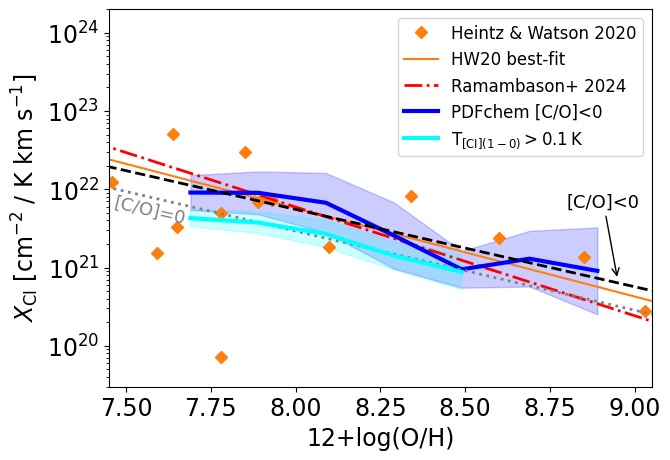}
    \caption{The $X_{\rm CO}$-factor (left panel) and the $X_{\rm CI}$-factor (right panel) for all models explored in this work (blue solid lines). The dashed black lines illustrate the best-fit expressions (Eqns~\ref{eqn:Xco} and \ref{eqn:Xci} for the $X_{\rm CO}$ and $X_{\rm CI}$, respectively). On the left panel, the observations of \citet{Israel97,Leroy11,Elmegreen13,Shi15,Shi16} are shown, including the average MW value marked in large blue star \citep{Bolatto13} and the best-fit relations of \citet{Genzel12} and \citet{Madden20}. On the right panel, the observations of \citet{Heintz20} and their best-fit relation is shown, as well as the best-fit relation of \citet{Ramambason24}. For all scattered observational data we plot the reported error bars, which may be smaller than the symbol size. In all cases, the {\sc PDFchem} best-fit expressions match well with observations. In both panels, the case of $\rm [C/O]=0$ is shown in gray dotted line which further assumes the linear relation of $\cal D$ with $\rm [O/H]$ (dotted lines in the panels of Fig.~\ref{fig:ics}). The cyan lines remark the conversion factors when we account for an observational sensitivity limit ($0.1\,{\rm K}$ for both CO~(1-0) and [C{\sc i}]~(1-0) lines). The shaded areas in the {\sc PDFchem} models represent $1\sigma$ standard deviation.}
    \label{fig:xfac}
\end{figure*}

Figure~\ref{fig:fixed} shows the behavior of both CO and C $X$-factors as a function of the FUV intensity for a fixed $\zeta_{\rm CR}=10^{-17}\,{\rm s}^{-1}$ cosmic-ray ionization rate (top panel) and as a function of the $\zeta_{\rm CR}$ for a fixed $\chi/\chi_0=1$ FUV intensity (bottom panel). Both panels refer to $\rm [O/H]=0$. As can be seen from the top panel, the FUV intensity does not affect the $X_{\rm CO}$-factor -which is in remarkable agreement with the MW average \citep[$X_{\rm CO}\sim2\times10^{20}\,{\rm cm}^{-2}\,({\rm K}\,{\rm km}\,{\rm s}^{-1})^{-1}$][]{Bolatto13}- regardless to the $A_{\rm V}$-PDF. However, as the $\chi/\chi_0$ increases, the $X_{\rm CI}$ decreases since the abundance of C, as well as the gas temperature increases leading to a brighter emission of the [C{\sc i}]~(1-0) line. The increase of the cosmic-ray ionization rate for fixed $\chi/\chi_0$ as illustrated in the bottom panel, causes a non-linear variation of the $X_{\rm CO}$-factor. Depending on the $m$, the $X_{\rm CO}$-factor slightly decreases for $\zeta_{\rm CR}=10^{-16}\,{\rm s}^{-1}$ before the formation of a local maximum at $\zeta_{\rm CR}\simeq10^{-14}\,{\rm s}^{-1}$ \citep[see also][]{Bell06}. This non-linear behavior is connected with an increase of the gas temperature at high column densities due to cosmic-ray heating, leading to a change of the CO formation pathway via the OH channel \citep{Bisbas17}. On the other hand, the $X_{\rm CI}$ factor shows a more linear and decreasing trend.

\subsection{General behavior of X-factors}
\label{ssec:general}

Figure~\ref{fig:xfac} shows how the $X_{\rm CO}$ and $X_{\rm CI}$ factors change as a function of metallicity, the latter written in the form $Z = {\rm 12+\log_{10}(O/H)}$. For all assumed $A_{\rm V}$ distributions, we consider the PDF averaged value of the conversion factor of each coolant. In these calculations, we exclude the very low and very high values of FUV intensities ($\chi/\chi_0<0.5$ and $\chi/\chi_0>10^2$) and high values of cosmic-ray ionization rates ($\zeta_{\rm CR}>10^{-15}\,{\rm s}^{-1}$), as all these represent more special systems. We also note that we have assumed equal weights to calculate the averages of all combinations considered, including the ISM environmental parameters and the inputted $A_{\rm V}$-PDF. 

The results of our models are shown in blue solid lines in both panels of Fig.~\ref{fig:xfac}. Black dashed lines refer to the best-fit values of the {\sc PDFchem} results for $\rm [C/O]<0$, which are given by the following expressions; for the CO-to-H$_2$ conversion factor
\begin{eqnarray}
    \label{eqn:Xco}
    \log_{10} X_{\rm CO} = -2.406Z+41.272
\end{eqnarray}
and for the [C{\sc i}]-to-H$_2$ conversion factor
\begin{eqnarray}
    \label{eqn:Xci}
    \log_{10} X_{\rm CI} = -0.989Z+29.653.
\end{eqnarray}
Our results are in broad agreement with observations which are described below. For comparison, we also plot the corresponding $X_{\rm CO}$- and $X_{\rm CI}$-factors for the linear decrease versus metallicity (marked as `$\rm [C/O]=0$') in dotted gray lines. Assuming a linear decrease of C and O abundances, and the dust-to-gas mass ratio, Eqns.~(\ref{eqn:Xco}) and (\ref{eqn:Xci}) take the form $\log_{10} X_{\rm CO,lin} = -1.869Z+36.681$ and $\log_{10} X_{\rm CI,lin} = -1.011Z+29.554$, respectively. As can be seen, $\rm [C/O]<0$ ratios tend to increase more the $X_{\rm CO}$-factor towards low metallicities comparing to $\rm [C/O]=0$, in better agreement with observations. Similarly, negative [C/O] tend to shift upwards the $X_{\rm CI}$-factor in the metallicity regime we explore.
Overall, we find that both $X$-factors show a declining trend with increasing [O/H], leading to sub-MW values for metal-rich ISM ($\rm [O/H]>0$). This is in line with several observations and models of metal-poor ISM using CO as a molecular gas tracer \citep[e.g.][]{Genzel12,Schruba12,Bolatto13,Amorin16, Tacconi18} or [C{\sc i}] \citep[e.g.][]{Accurso17,Madden20,Bisbas21}. 

The observational data shown in Fig.~\ref{fig:xfac} are as follows. In the left panel, the large blue star marks the average $X_{\rm CO}$-factor value of our Galaxy \citep{Bolatto13}. 
Green crosses correspond to the SMC and LMC observations \citep{Israel97}\footnote{There are two values of $X_{\rm CO}$ for NGC~4499 in their Table~3 and 4. We have used the data of Table~4 with the error scaled accordingly.}. Red diamonds correspond to a collection of observations from the Local Group of galaxies \citep{Leroy11}. 
With orange triangle, we plot the observed value of the WLM galaxy \citep{Elmegreen13}, which is a dwarf irregular galaxy with metallicity $Z=0.13\,{\rm Z}_{\odot}$ observed in CO~(3-2). The purple square shows the $X_{\rm CO}$-factor of the very metal poor ($Z\sim0.07\,{\rm Z}_{\odot}$) galaxy Sextans~A \citep{Shi15}. With brown-filled circles we plot the \citet{Shi16} CO-to-H$_2$ conversion factor of the very metal-poor dwarf galaxy DDO-70 as well as of DDO-53 and DDO-50. All aforementioned works have estimated the H$_2$ column density from dust emission, while \citet{Elmegreen13} have further assumed a CO~(3-2)/CO~(1-0)$=0.8$ ratio to convert the observed CO~(3-2) to CO~(1-0). Furthermore, we plot two best-fit relations from the literature. With thin dashed black line the one presented in \citet{Genzel12} resulting from a sample of high-redshift ($z\ge1$) star-forming galaxies of the so-called `main-sequence' in which the H$_2$ gas mass has been estimated from the Star-Formation Rate using the Kennicutt-Schmidt relation \citep{Kennicutt98}. With dotted red line we plot the one presented in \citet{Madden20} resulting from the Dwarf Galaxy Survey \citep[`DGS'][]{Madden13} who used {\sc Cloudy} models to estimate the H$_2$ gas mass from [C{\sc ii}] and CO~(1-0) observations.

The aforementioned curves resulting from the {\sc PDFchem} algorithm (Eqns.~\ref{eqn:Xco} and \ref{eqn:Xci}), consider the total emission and total H$_2$ column density of each distribution without adopting any limit concerning observational sensitivities. In reality, it can be expected that the faintest part of the CO emission will not be captured by radio telescopes, leaving a portion of H$_2$-rich gas undetected using this line and making it `CO-dark', as discussed also below. To accommodate this, we plot the CO-to-H$_2$ conversion factor when account is taken of such an observational sensitivity. In particular, we adopt a minimum brightness temperature of $T_{\rm CO~(1-0)}=0.1\,{\rm K}$ which is representative of the typical sensitivity limit in observations \citep[e.g.][]{Nieten06, Smith14, Leroy16, Tokuda21, Luo24}, and we also calculate the H$_2$ column density for $T_{\rm CO(1-0)}\ge0.1\,{\rm K}$. We then calculate both the PDF-averaged brightness temperature of the line, and the PDF-averaged H$_2$ column density above the aforementioned limit. Their ratio, therefore, represents a CO-to-H$_2$ conversion factor that is tightly connected with the CO-bright H$_2$ gas, omitting the CO-dark component. As can be seen in the left panel of Fig.~\ref{fig:xfac}, the discrepancy between the blue and cyan lines is very small for solar metallicities as the CO-dark gas fraction is minor under such conditions (see discussion in \S\ref{ssec:codark}). As metallicity decreases, this discrepancy increases reaching approximately two orders of magnitude for the lowest metallicity case explored. This implies that in such low metallicities, the CO-bright component may account for just a few percentage of the real H$_2$ gas mass existing in these distributions. As can be expected, higher sensitivity limits tend to shift upwards the cyan CO-to-H$_2$ curve.

\begin{figure*}
    \centering
    \includegraphics[width=0.49\textwidth]{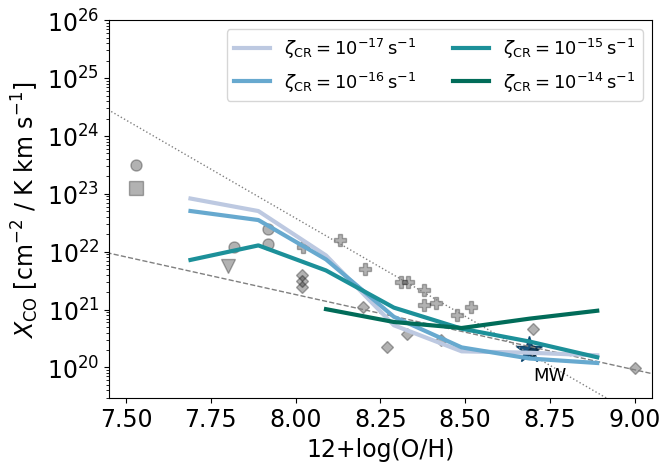}
    \includegraphics[width=0.49\textwidth]{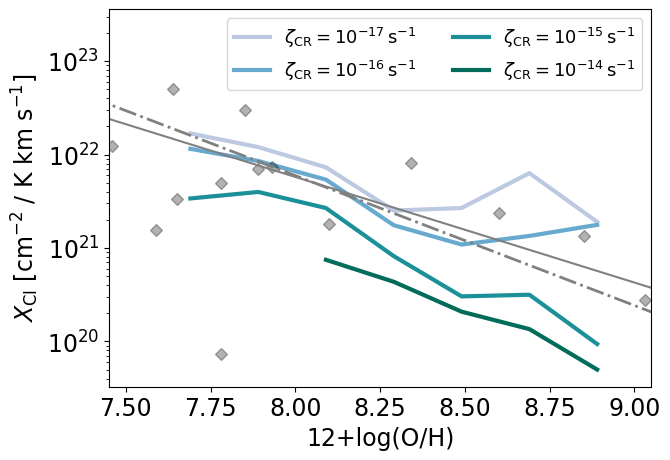}
    \includegraphics[width=0.49\textwidth]{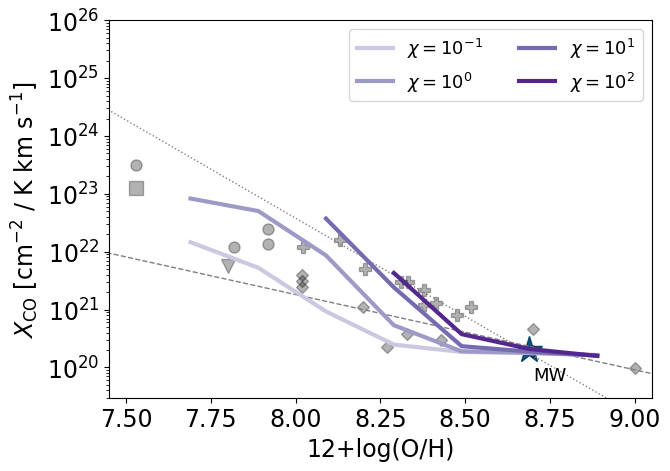}
    \includegraphics[width=0.49\textwidth]{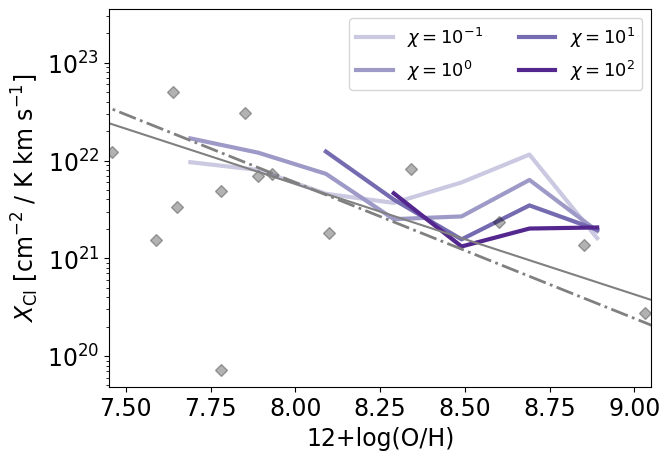}
    \includegraphics[width=0.49\textwidth]{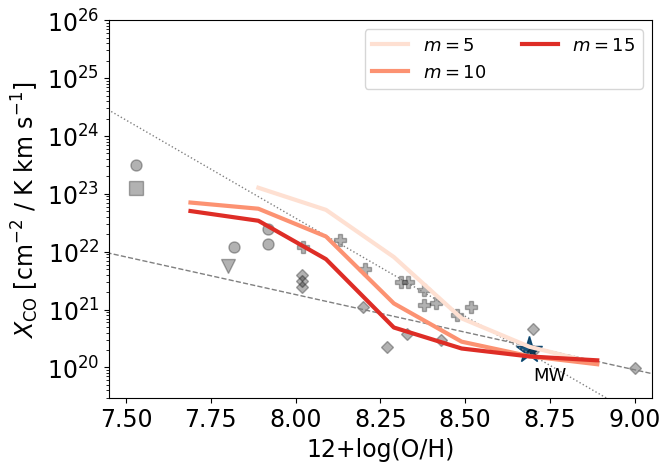}
    \includegraphics[width=0.49\textwidth]{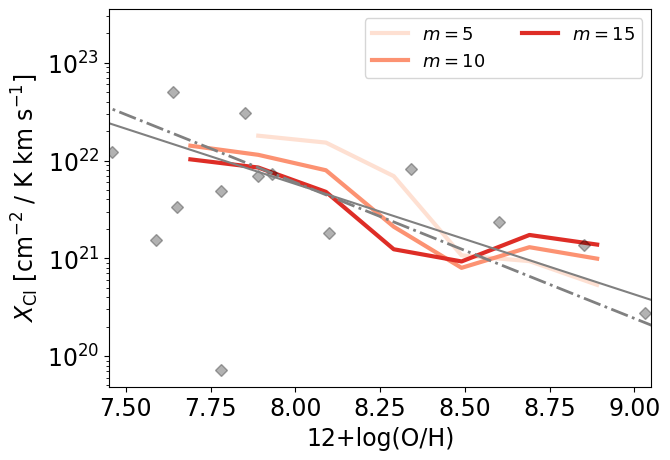}
    \caption{The $X_{\rm CO}$ (left column) and $X_{\rm CI}$ (right column) conversion factors versus metallicity, as a function of the FUV intensity (top row), the cosmic-ray ionization rate (middle row) and the density distribution (bottom row). Top row panels illustrate the dependency on $\zeta_{\rm CR}$ for a fixed FUV intensity of $\chi/\chi_0=1$. Middle row panels show the dependency on $\chi/\chi_0$ for a fixed $\zeta_{\rm CR}=10^{-17}\,{\rm s}^{-1}$. In both top and middle rows we consider all three $m$ and three $\sigma$ values (nine distributions in total) of Table~\ref{tab:avpdfs}. Bottom row panels illustrate the dependency on the mean, $m$, of the $A_{\rm V}$-PDF for a fixed $\sigma_{\rm PDF}=0.4$. The bottom panels are averages for $10^{-17}\le\zeta_{\rm CR}\le10^{-15}\,{\rm s}^{-1}$ and $0.5\le\chi/\chi_0\le10^2$. In all panels, the gray scatter points and the gray lines correspond to the observations and best-fit relations described in Fig.~\ref{fig:xfac}.}
    \label{fig:xfacdep}
\end{figure*}

In the right panel of Fig.~\ref{fig:xfac}, the observations of \citet{Heintz20} are shown along with their best-fit relation. These are high-redshift observations of star-forming galaxies using gamma-ray burst and quasar absorbers to estimate the H$_2$ column density and the [C{\sc i}] luminosity. We also plot the best-fit relation of \citet{Ramambason24} in dot-dashed red line, corresponding to low-metallicity dwarf galaxies from the DGS. In their study, \citet{Ramambason24} used the statistical algorithm {\sc multigris} and the PDR code {\sc Cloudy} \citep{Ferland17} to estimate the luminosity of [C{\sc i}] through CO lines. 
As with the CO-to-H$_2$ case, we also account here for a similar observational sensitivity of $T_{\rm [CI]~(1-0)}=0.1\,{\rm K}$. The cyan line illustrated in the right panel of Fig.~\ref{fig:xfac}, indicates that the [C{\sc i}]~(1-0) fine-structure line remains a more robust tracer of molecular gas even at very low metallicities. Here, we should mention also that this line may not always arise \emph{directly} from the molecular gas component but rather from the PDR surface for typical ISM conditions concerning the FUV and $\zeta_{\rm CR}$ values. For high $\zeta_{\rm CR}$ however, the [C{\sc i}]~(1-0) line originates directly from the H$_2$ gas as has been discussed in previous works \citep{Bisbas15,Bisbas17,Bisbas21}.

\begin{figure*}
    \centering
    \includegraphics[width=0.95\textwidth]{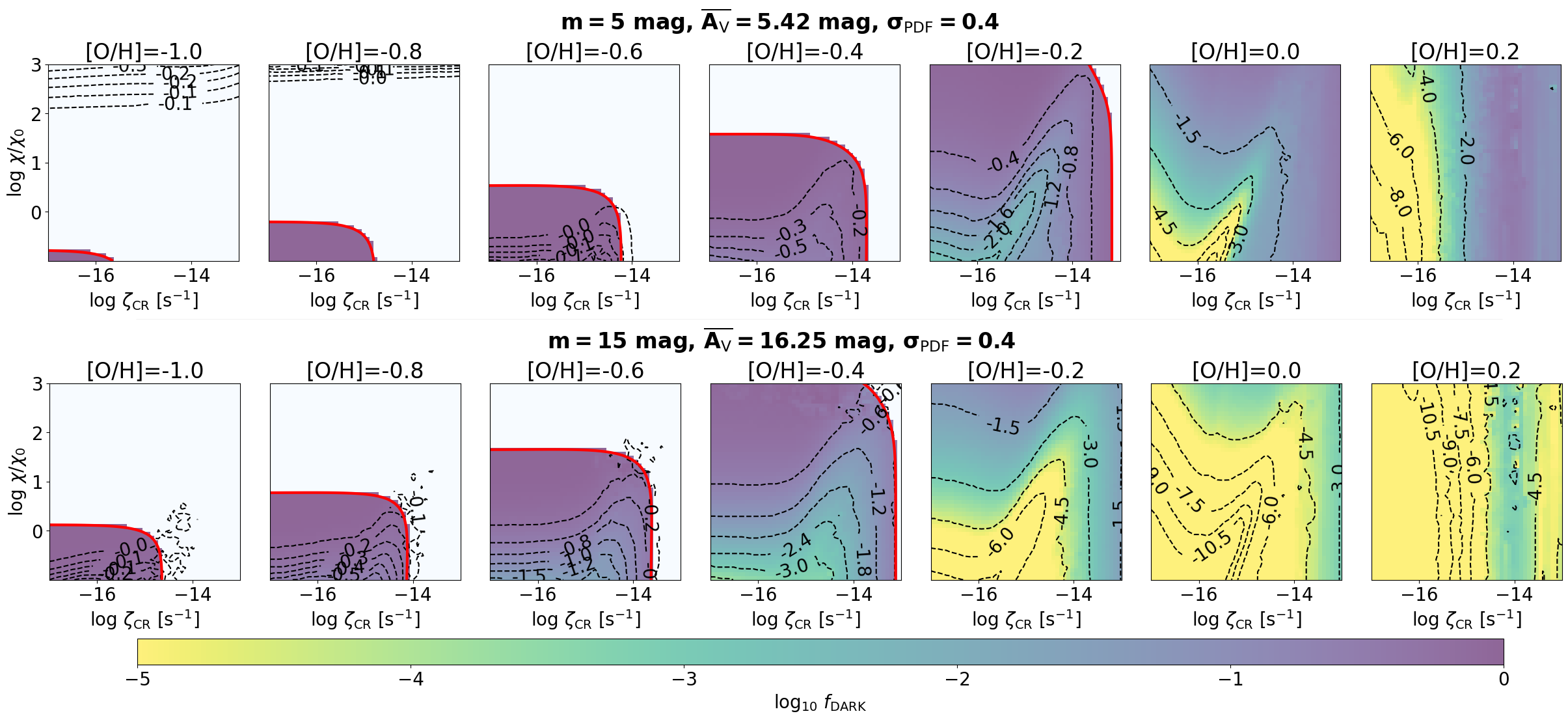}
    \caption{Logarithmic grid maps as in shown in Figs.~\ref{fig:m5} and \ref{fig:m15} but colored with respect to the logarithmic $f_{\rm DARK}$ CO-dark gas fraction. The upper row shows results for the $m=5$ and $\sigma_{\rm PDF}=0.4$ distribution, and the bottom row for the $m=15$ and $\sigma_{\rm PDF}=0.4$ distribution. In general, we find that the H$_2$-rich gas becomes progressively CO-dark as [O/H] decreases. For $\rm [O/H]\lesssim-0.4$, $f_{\rm DARK}$ depends stronger on the FUV intensity than the CR ionization rate. It is expected that $A_{\rm V}$-PDFs with lower $m$ than those explored here (e.g. representing a diffuse cloud) will have higher $f_{\rm DARK}$ fractions for $\rm [O/H]\gtrsim0$.}
    \label{fig:codark}
\end{figure*}

Although Fig.~\ref{fig:xfac} shows already how both conversion factors depend -on average- on metallicity, it would be interesting to explore how individually the cosmic-ray ionization rate, the FUV intensity and the density distribution affect both $X_{\rm CO}$ and $X_{\rm CI}$ as a function of metallicity. 

\subsubsection{Dependence on the cosmic-ray ionization rate}

The top row of Fig.~\ref{fig:xfacdep} shows how increasing the $\zeta_{\rm CR}$ parameter changes the conversion factors, by keeping the FUV radiation fixed at $\chi/\chi_0=1$. We find that for $\zeta_{\rm CR}\le10^{-16}\,{\rm s}^{-1}$, the $X_{\rm CO}$ (left panel) remains approximately constant for metallicities down to $12+\log_{10}(\rm O/H)\simeq8.4$. For lower metallicities, the conversion factor increases approaching the \citet{Madden20} best-fit. With respect to the higher CR values, for $\zeta_{\rm CR}=10^{-15}\,{\rm s}^{-1}$ the $X_{\rm CO}$-factor increases but not as strongly as with lower $\zeta_{\rm CR}$, because the average gas temperature increases making the CO~(1-0) brighter in high density regions \citep{Bisbas21} mitigating the $X_{\rm CO}$ value. For $\zeta_{\rm CR}=10^{-14}\,{\rm s}^{-1}$ it remains approximately constant at $X_{\rm CO}\sim10^{21}\,{\rm cm}^{-2}/{\rm K}\,{\rm km}\,{\rm s}^{-1}$ for all metallicities. Note that for that high $\zeta_{\rm CR}$ and $12+\log_{10}(\rm O/H)\lesssim8.1$, none of the assumed $A_{\rm V}$-PDFs is H$_2$-dominated. On the other hand, the $X_{\rm CI}$-factor appears to depend stronger on the $\zeta_{\rm CR}$ value, although the dispersion in $X_{\rm CI}$ values is much smaller than that of $X_{\rm CO}$ (note the smaller extent of values in the $y$-axis). Interestingly, for each cosmic-ray ionization rate, the [C{\sc i}]-to-H$_2$ conversion factor remains to within a factor of $\sim5$ in low-metallicity as opposed to the CO-to-H$_2$ which may change by more than two orders of magnitude. For solar metallicities, however, we find that $X_{\rm CI}$ does depend strongly on the $\zeta_{\rm CR}$ rate with the value of the conversion factor to span approximately two orders of magnitude. This spread declines as metallicity decreases. In this regard, for fixed $\chi/\chi_0$ and $\zeta_{\rm CR}$, $X_{\rm CI}$ does not change significantly in low-$Z$. 

\subsubsection{Dependence on the FUV intensity}

The middle row of Fig.~\ref{fig:xfacdep} shows the dependency on the strength of the FUV radiation field for a fixed $\zeta_{\rm CR}=10^{-17}\,{\rm s}^{-1}$. It appears that for solar metallicities and above, the $X_{\rm CO}$-factor does not depend on the value of $\chi/\chi_0$ for the collection of the $A_{\rm V}$-PDF explored (see also top panel of Fig.~\ref{fig:fixed}). However, as metallicity decreases, low FUV intensities tend to return a lower CO-to-H$_2$ conversion factor than higher FUV intensities. This is because as the radiation strength increases, photodissocation changes the depth point where the transition phase of the carbon cycle (C$^+$/C/CO) occurs, leading to a slower build in CO abundance and thus a weaker CO~(1-0) emission \citep[see also][]{Bell06}. It is interesting to note that the \citet{Israel97} estimations of the $X_{\rm CO}$-factor are in a better agreement with strong FUV radiation fields than with cosmic-ray ionization rates. Indeed, stronger FUV intensities than the average MW in the diffuse ISM of the Magellanic Clouds have been reported \citep{Pradhan11,Israel11}. Note also that for high $\chi/\chi_0$, the density distributions may not be H$_2$-rich due to photodissociation. On the other hand, the $X_{\rm CI}$-factors shown on the right panel of the middle row do not depend as strongly on the FUV intensity across metallicities as the $X_{\rm CO}$ does. However, for solar metallicities (e.g. the MW), the [C{\sc i}]-to-H$_2$ factor has a spread of approximately one order of magnitude; lower $\chi/\chi_0$ are connected with higher $X_{\rm CI}$ values. This trend appears to reverse in low-metallicities. Interestingly, a variation on the FUV intensities cannot reproduce the observed [C{\sc i}]-to-H$_2$ conversion factors below the best-fit functions of \citet{Heintz20} and \citet{Ramambason24}. 

\subsubsection{Dependence on the column-density distribution}

In the bottom row of Fig.~\ref{fig:xfacdep}, we plot the dependency of the conversion factors on the mean ($m$) of the $A_{\rm V}$-PDFs used in this work. For these results, we assumed a fixed width of $\sigma_{\rm PDF}=0.4$ and we have considered FUV intensities and cosmic-ray ionization rates in the ranges of $0.5\le\chi/\chi_0\le10^2$ and $10^{-17}\le\zeta_{\rm CR}\le10^{-15}\,{\rm s}^{-1}$, respectively. As can be seen, both conversion factors depend on the value of $m$; lower mean values produce in general a larger $X$-factor. It is interesting to note that the SMC and LMC observations \citep{Israel97} agree with low-density distributions as well. A dependency on the column density distribution may be connected with the evolutionary stage of clouds as well; such a relation has been previously observed in the low-metallicity dwarf galaxy NGC~6822 \citep{Schruba17} with CO-to-H$_2$ conversion factors exceeding $\sim20\times$ the MW value. The $X_{\rm CI}$-factor shows a much weaker dependency on the column density distribution. In agreement with these findings are the three dimensional PDR models of \citet{Bisbas21} that examine the conversion factors of a star-forming and a non-star-forming density distributions, who find that high density structures tend to have a much lower $X_{\rm CO}$-factor in metal poor ISM than low density ones, whereas their $X_{\rm CI}$-factors remain approximately constant, even for supersolar metallicities.

\subsection{CO-dark molecular gas fraction}
\label{ssec:codark}

As described in the \nameref{sec:intro}, `CO-dark' refers to the cold H$_2$-rich gas that is not observed in CO line emission. There are essentially two reasons resulting in a CO-dark molecular gas; i) the absence of CO molecules in molecular regions existing in translucent and diffuse clouds due to the weaker CO self-shielding than that of H$_2$, and ii) the sensitivity limit of telescopes which are able to detect the CO~(1-0) line emission above some threshold. In the observational context, the term `CO-dark' refers to the second of the above cases, while the first one is better studied (if not \emph{only} studied) with numerical modeling \citep[see also discussion in][]{Seifried20}.

In this work, we adopt the observational definition of CO-dark, that is the H$_2$-rich gas which has a CO~(1-0) brightness temperature of $<0.1\,{\rm K}$ (chosen limit as discussed above). In turn, the CO-dark gas fraction, $f_{\rm DARK}$, is defined here as the ratio between the H$_2$ column density that has a $T_{\rm CO(1-0)}<0.1\,{\rm K}$ to the total H$_2$ column density of the PDR model:
\begin{eqnarray}
    f_{\rm DARK} = \frac{N({\rm H}_2)_{\rm CO-dark}}{N({\rm H}_2)_{\rm total}}.
\end{eqnarray}

\begin{figure}
    \centering
    \includegraphics[width=0.4\textwidth]{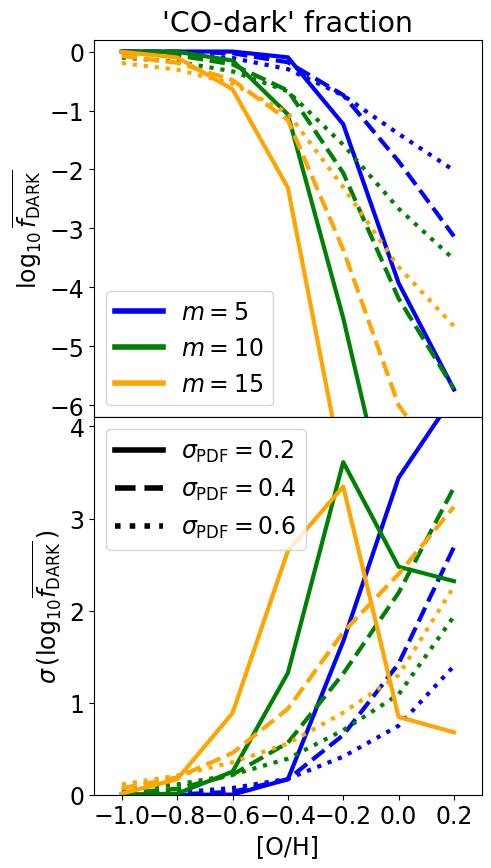}
    \caption{Mean values (top) of the logarithmic $X$ conversion factor for all $A_{\rm V}$-PDFs considered in this work, and the corresponding standard deviation around the mean (bottom). \textit{Top panel:} The average, $\log_{10}\overline{f_{\rm DARK}}$, fraction for all $\zeta_{\rm CR}-\chi/\chi_0$ pairs at constant [O/H] and $A_{\rm V}$-PDF properties. We find that $f_{\rm DARK}$ increases with decreasing metallicity making the gas severely CO-dark, especially for $\rm [O/H]\sim-1.0$. \textit{Bottom panel:} The ambient ISM environmental parameters play an important role in determining the $f_{\rm DARK}$, particularly for the higher [O/H] ratios and the lower $\sigma_{\rm PDF}$ widths. The large drops of $\sigma$ in the $m=10,15$ at $\sigma=0.2$ (orange and green solid lines) are because for these distributions, a nearly uniform value of $f_{\rm DARK}$ is obtained for any pair of $\chi/\chi_0-\zeta_{\rm CR}$.}
    \label{fig:codark_collective}
\end{figure}

Figure~\ref{fig:codark} shows logarithmic grid maps of the $f_{\rm DARK}$ ratio for the two representative $A_{\rm V}$-PDFs discussed in \S\ref{ssec:detailed}. In both cases -implying that a similar result should be expected for all distributions considered- we find that the H$_2$-rich gas is severely CO-dark for $\rm [O/H]=-1.0$. It is only for metallicities $\rm [O/H]\ge-0.8$ and for denser ($m=15\,{\rm mag}$) distributions that the molecular gas will be CO-bright, and that for very low FUV intensities. As metallicity increases towards $\rm [O/H]=0$ and above, the fraction of CO-dark gas decreases overall. As expected, for higher $m$ values, the effect of a decreasing $f_{\rm DARK}$ is enhanced. In addition, for a constant [O/H], $f_{\rm DARK}$ increases with increasing $\zeta_{\rm CR}$ when $\zeta_{\rm CR}\gtrsim10^{-15}\,{\rm s}^{-1}$ due to the CR-induced destruction of CO. For lower $\zeta_{\rm CR}$ the CO-dark gas fraction depends primary on the FUV intensity.

Figure~\ref{fig:codark_collective} shows the general trends of the CO-dark gas for all $A_{\rm V}$-PDFs considered here. For each distribution, the plotted curves represent the average of the molecular region (colored partition in Fig.~\ref{fig:codark}). From the top panel we find that $f_{\rm DARK}$ increases with decreasing [O/H]. As expected, the denser the molecular cloud is (higher $m$), the lower the $f_{\rm DARK}$ is for fixed [O/H] when compared to a more diffuse molecular cloud (lower $m$). We also find that for fixed $m$, the width ($\sigma_{\rm PDF}$) of the distribution does not considerably change the CO-dark fraction. In regards to the influence of the ambient ISM environmental parameters ($\zeta_{\rm CR}$ and $\chi/\chi_0$) on $f_{\rm DARK}$, from the lower panel of Fig.~\ref{fig:codark_collective} we find that the standard deviation, $\sigma(\log_{10}\overline{f_{\rm DARK}})$, peaks for $\rm [O/H]\sim-0.4$ to $-0.2$. We therefore argue that for $\rm [O/H]\sim-0.2$ (the average metallicity of the ISM in the solar neighbourhood), both $\zeta_{\rm CR}$ and $\chi/\chi_0$ play an important role in estimating the H$_2$ gas mass through the commonly used CO~(1-0) emission line. 

\begin{figure*}
    \centering
    \includegraphics[width=0.98\textwidth]{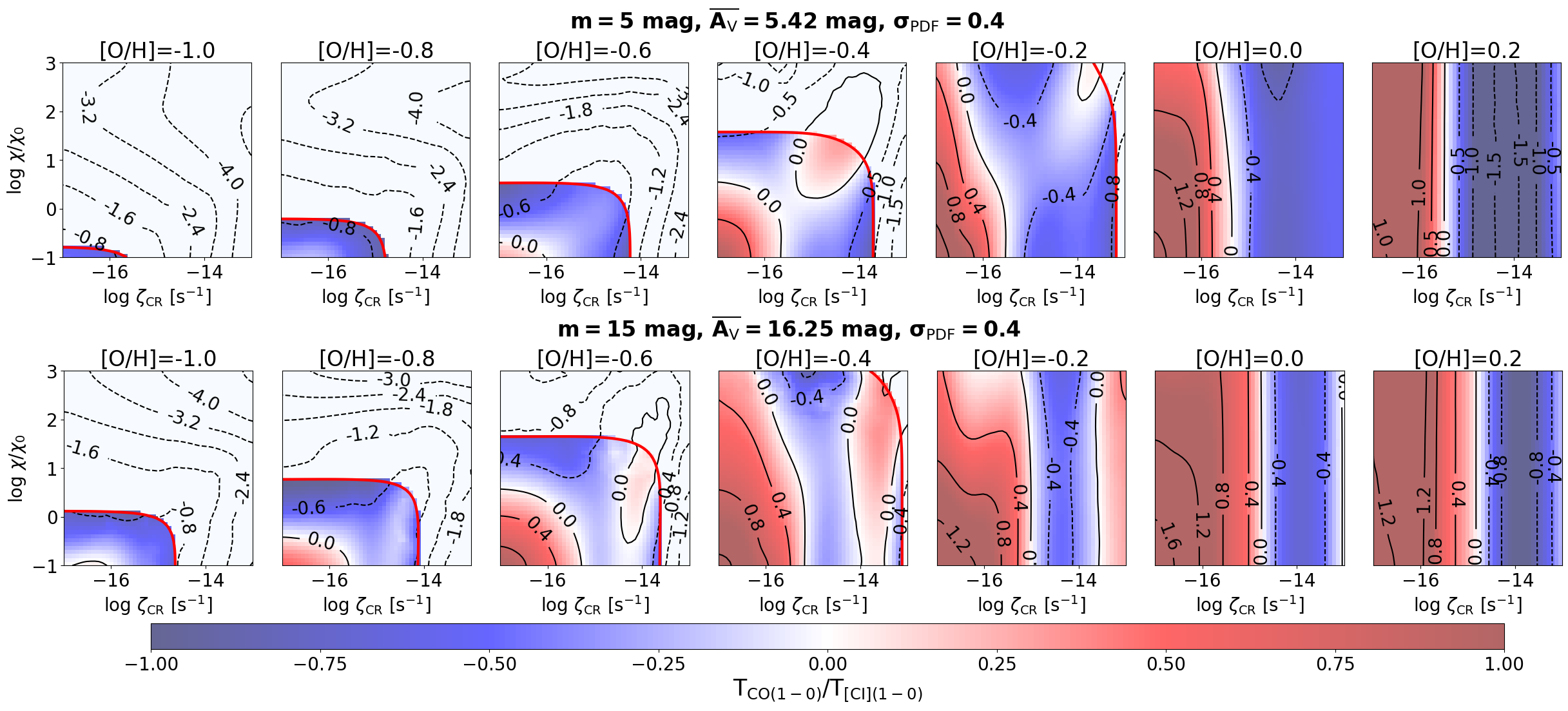}
    \caption{Logarithmic grid maps showing the ratio of CO~(1-0) and [C{\sc i}]~(1-0) brightness temperatures. The color bar corresponds to the logarithm of that ratio. Positive values (red colored) indicate a stronger CO~(1-0) brightness temperature whereas negative values (blue colored) a stronger [C{\sc i}]~(1-0) brightness temperature. The upper row corresponds to the $m=5$ and $\sigma_{\rm PDF}=0.4$, while the bottom row to the $m=15$ and $\sigma_{\rm PDF}$ distribution.}
    \label{fig:ratios}
\end{figure*}

While there exists a large fraction of CO-dark molecular gas particularly in the low metallicity ISM as it was found above, it would be interesting to see whether this H$_2$ gas can be observed using the [C{\sc i}]~(1-0) fine-structure line instead. Figure~\ref{fig:ratios} shows how the brightness temperature ratio of CO~(1-0) and [C{\sc i}]~(1-0) change as a function of $\zeta_{\rm CR}$ and $\chi/\chi_0$ for the two representative $A_{\rm V}$-PDFs discussed in \S\ref{ssec:detailed}. For metallicities close to solar (e.g. $\rm [O/H]\gtrsim-0.2$) and above, and for low cosmic-ray ionization rates (e.g. $\zeta_{\rm CR}\lesssim10^{-15}\,{\rm s}^{-1}$), we find that the brightness temperature of CO~(1-0) is much higher than that of [C{\sc i}]~(1-0). The [C{\sc i}]~(1-0) becomes brighter for higher $\zeta_{\rm CR}$ values, in agreement with the earlier findings of \citet{Bisbas15,Bisbas17,Bisbas21}. For lower metallicities, we find that the [C{\sc i}]~(1-0) line dominates, except for combinations of low $\zeta_{\rm CR}\lesssim10^{-16}\,{\rm s}^{-1}$ and low $\chi/\chi_0\lesssim1$. Distributions reflecting high-density/star-forming gas (e.g. $m=15$) may contain small clumps of high depths in column that remain bright in CO~(1-0) even for very low metallicities \citep{Madden20} provided that cosmic-rays and the intensity of the FUV radiation have been severely attenuated. We, thus, find that the [C{\sc i}]~(1-0) line may be a good alternative line for tracing molecular gas in environments of low metallicity. The much smaller variation of the $X_{\rm CI}$ conversion factor compared to $X_{\rm CO}$ versus metallicity (Fig.~\ref{fig:xfac}) also supports this result.

\section{Discussion}
\label{sec:discussion}


From the results of the models presented both in \citet{Bisbas24} and here, it is evident that a non-linear C/O ratio versus metallicity, leading to an $\alpha$-enhanced ISM, has important implications in studying metal-poor galaxies. In general, we find a strong influence in the behavior of molecular gas tracers and the thermal balance of the ISM. This influence arises primary due to the much less efficient dust-shielding in low-metallicity environments -leading also to a much less efficiency in H$_2$ formation- and secondary due to the lower C/O ratios which affect the total gas cooling.

In low-metallicity environments, the C/O ratio is affected by the distinct stellar nucleosynthesis pathways of carbon and oxygen, causing a differential enrichment of C and O as a function of O/H. This leads to changes in the efficiency of molecular gas tracers. In particular, we find that as CO becomes much less abundant in low-metallicity environments due to the reduced self-shielding and lower formation efficiency, the [C{\sc i}] emission may become more prominent as it traces the regions where CO formation is less efficient. Since the non-linear C/O ratio as well as the dust-to-gas ratios affect the thermal balance of the ISM, it is expected that in low-metallicity galaxies the physical conditions of the ISM influencing the efficiency of the star formation will be different affecting the initial mass function \citep[e.g.][]{Marks12,Bate19,Sharda22,Tanvir24}. 

However, for metallicities higher than the solar one, we find that the conversion factors become very stable against most of the ISM environmental parameters we explored. For $\rm [O/H]=0.2$, our models show that the FUV radiation field and the column-density distribution of the cloud (represented by the considered $A_{\rm V}$-PDFs) do not affect either the CO-to-H$_2$ or the [C{\sc i}]-to-H$_2$ conversion factors. It is only when a higher cosmic-ray ionization rate is applied that the $X$-factors of both these tracers change. Notably, for higher $\zeta_{\rm CR}$ values the $X_{\rm CO}$ increases while $X_{\rm CI}$ decreases. This is connected to the efficient destruction of CO by cosmic-rays leading to a more extended [C{\sc i}]-bright molecular gas \citep{Bisbas15,Bisbas21}. For such supersolar metallicities, our models show that effects connected to different C/O and dust-to-gas mass ratios from the commonly used linear assumptions, cease to be important.

Our models assumed the density function arising from the $A_{\rm V,eff}-n_{\rm H}$ relation and used this for all thermochemical models presented here. 
This relation has been found to operate in small (pc) and large (kpc) scale structures \citep[e.g.][]{Glover10,VanLoo13,Safranek17,Seifried17,Hu21}.
In our models, we assume that the density distribution does not change with metallicity, an assumption that has been made in previous similar works \citep[e.g.][]{Bisbas19,Bisbas23,Bisbas24}. However, in low metallicity environments the weaker shielding due to dust in combination with the lower abundances of metals changes the thermal balance impacting the overall density distribution. Very few models exploring this $A_{\rm V,eff}-n_{\rm H}$ relationship and its connection to metallicity variations have been reported. The hydrodynamical models of \citet{Hu21} offer such an insight and find a correlation of the form $A_{\rm V,eff}\propto n_{\rm H}^{\alpha}$, where $\alpha$ increases slightly as metallicity decreases. This indicates a systematic shift in the relationship at lower metallicities due to changes in dust shielding and cloud structure. In general, \citet{Hu21} do not report a strong dependency of that relation with metallicity, making our results robust.

In regards to the CO-dark fraction of molecular gas, we find that it highly increases at low metallicities. In addition, diffuse clouds contain higher CO-dark fractions than denser / star-forming clouds even at solar metallicities. These trends are in agreement with observational works previously reported. For instance, earlier {\it Herschel} studies by \citet{Langer14} on individual clouds within MW found that the fraction of CO-dark H$_2$ gas decreases with increasing total column density, and that clouds in the outer/metal-poor regions of our Galaxy have greater fraction of CO-dark H$_2$, consistent with our results. In line with a larger fraction of CO-dark H$_2$ gas in low-metallicity gas is the study of \citet{Balashev17} in a high-redshift ($z\simeq2.786$) damped Ly~$\alpha$ absorption system, which is found to contain large amounts of H$_2$ gas but poor CO-bright gas, evident from the very high $1-3\times10^3$ times the Galactic CO-to-H$_2$ conversion factor. More recently, \citet{Chevance20} reported a large fraction of CO-dark gas in the massive star-forming region 30~Doradus in the LMC, which has a half-solar metallicity. 

Our finding that the $X_{\rm CO}$-factor exhibits a larger dispersion as metallicity decreases is consistent with observational studies. For instance, \citet{Leroy11} and \citet{Schruba12} have reported significant variations in the $X_{\rm CO}$-factor across low-metallicity environments, with values ranging from a few times to several hundred times the Galactic average. These variations are attributed to the complex relationship between metallicity, dust shielding, and the local ISM conditions. Similarly, \citet{Madden20} observed a wide range of $X_{\rm CO}$ values in low-metallicity dwarf galaxies, further supporting the idea that the CO-to-H$_2$ conversion factor becomes increasingly uncertain in metal-poor environments. In line with this trend of a larger dispersion as metallicity decreases, are the observations reported by \citet{Elmegreen13} and \citet{Shi15,Shi16}, as illustrated in the left panel of Fig.~\ref{fig:xfac}.

The [C{\sc i}]-to-H$_2$ factor appears to be a reliable choice for molecular gas studies in low-metallicity environments. Our models show that the $X_{\rm CI}$ factor shows a small variability with changes in ISM parameters, such as the FUV intensity and the cosmic-ray ionization rate. This variability is also found to be less than the corresponding one for the $X_{\rm CO}$-factor. While both $X_{\rm CI}$ and $X_{\rm CO}$ increase as metallicity decreases, the CO-to-H$_2$ conversion factor exhibits a much steeper rise, making it more difficult to interpret in extreme environments with low [O/H]. In contrast, the response of $X_{\rm CI}$ in such changes is more gradual and predictable. A significant fraction of molecular gas in low metallicities cannot be effectively traced by the CO emission \citep{Bolatto13,Chevance20,Madden20}. We find that the [C{\sc i}]~(1-0) line is closely linked to the H$_2$-rich and CO-deficit regions, thus better suited for the detection of this molecular gas component.

\section{Conclusions}
\label{sec:conclusions}

This work continues the efforts of \citet{Bisbas24} in studying the impact of $\alpha$-enhanced ISM environments on observables. In particular, we studied the impact of negative [C/O] ratios and observationally-driven dust-to-gas ratios versus metallicity, in the CO-to-H$_2$ and the [C{\sc i}]-to-H$_2$ conversion factors. Our study used the recently developed algorithm, {\sc PDFchem}, which estimates the average PDR quantities from entire distributions of column densities, using a set of pre-calculated 1D PDR models following the $A_{\rm V,eff}-n_{\rm H}$ relation modeled with {\sc 3d-pdr}. We have also considered a set of $A_{\rm V}$-PDF functions interacting with a wide range of FUV intensities and cosmic-ray ionization rates, covering many different combinations that represent quiescent and extreme conditions. 

Our study finds that the CO-to-H$_2$ conversion factor strongly depends on metallicity, and that it increases significantly as metallicity decreases. For environments with $\rm [O/H]=-1.0$, $X_{\rm CO}$ is $\gtrsim1000\times$ higher than the average Galactic value for $\rm [C/O]<0$ conditions, whereas it is $\lesssim300\times$ for $\rm [C/O]=0$, highlighting the significant sensitivity of the C/O and dust-to-gas ratios in studying the metal-poor ISM. The trend of increasing $X_{\rm CO}$ is consistent with the reduced CO formation efficiency in low-metallicity conditions. However, the CO-to-H$_2$ factor also exhibits sensitivity to ISM environmental parameters, particularly the cosmic-ray ionization rate, which can lead to highly non-linear variations. 

The [C{\sc i}]-to-H$_2$ conversion factor shows a moderate dependency on metallicity compared to $X_{\rm CO}$, and it decreases with increasing metallicity but within a narrower range. $X_{\rm CI}$ is less sensitive to variations in the FUV radiation field at low metallicities, making it a more reliable tracer of molecular gas in such environments. This reliability highlights its potential as an alternative to CO for molecular gas mass estimation in metal-poor galaxies.

The fraction of CO-dark molecular gas increases significantly in low-metallicity ISM, with values exceeding 90\% for $\rm [O/H] = -1.0$. This reflects the inefficiency of CO self-shielding in such environments, where H$_2$ can form and persist without detectable CO emission. The CO-dark fraction is strongly influenced by the FUV intensity and cosmic-ray ionization rate, particularly at intermediate metallicities ($\rm [O/H] \simeq -0.4$). These findings underline the challenges in accurately estimating molecular gas content using CO emission alone in metal-poor systems.

For estimating the molecular mass content in galaxies with metallicities in the range of $-1<\rm [O/H]<0.2$ examined here, we recommend the use of Eqns.~\ref{eqn:Xco} and \ref{eqn:Xci} for the CO-to-H$_2$ and [C{\sc i}]-to-H$_2$, respectively. However, we note that in very metal-poor conditions the conversion factors -especially for $X_{\rm CO}$- depend strongly on the local ISM environmental conditions with variations ranging to more than an order of magnitude. It is, therefore, imperative to examine each different system carefully and individually.

\begin{acknowledgements}
The authors thank the anonymous referee for their comments that help improve the clarity of our work.
This work is supported by the National Key Research \& Development (R\&D) Programme of China (Grant No. 2023YFA1608204).  
This work is supported by the Leading Innovation and Entrepreneurship Team of Zhejiang Province of China (Grant No. 2023R01008). 
Part of this work was supported by the German
\emph{Deut\-sche For\-schungs\-ge\-mein\-schaft, DFG\/} project
number Ts~17/2--1. Z.-Y.Z. acknowledges the support of the National Natural Science Foundation of China (NSFC; Grant Nos. 12173016 and 12041305), and the Programme for Innovative Talents, Enterpreneur in Jiangsu.
Y.Z. is grateful for support from the National Natural Science Foundation of China (NSFC) under grant No. 12173079.
T.T. gratefully acknowledges the Collaborative
Research Center 1601 (SFB 1601 subproject A6) funded by the Deutsche
Forschungsgemeinschaft (DFG, German Research Foundation) – 500700252. DL acknowledges support from the National Natural Science Foundation of China (NSFC) Grant No. 11988101. DL is a New Cornerstone Investigator. 
\end{acknowledgements}

\bibliographystyle{aa}
\bibliography{refs}

\begin{thebibliography}{148}
\expandafter\ifx\csname natexlab\endcsname\relax\def\natexlab#1{#1}\fi

\bibitem[{{Abreu-Vicente} {et~al.}(2015){Abreu-Vicente}, {Kainulainen},
  {Stutz}, {Henning}, \& {Beuther}}]{Abreu-Vicente15}
{Abreu-Vicente}, J., {Kainulainen}, J., {Stutz}, A., {Henning}, T., \&
  {Beuther}, H. 2015, \aap, 581, A74

\bibitem[{{Accurso} {et~al.}(2017){Accurso}, {Saintonge}, {Bisbas}, \&
  {Viti}}]{Accurso17}
{Accurso}, G., {Saintonge}, A., {Bisbas}, T.~G., \& {Viti}, S. 2017, \mnras,
  464, 3315

\bibitem[{{Aller} \& {Greenstein}(1960)}]{Aller60}
{Aller}, L.~H. \& {Greenstein}, J.~L. 1960, \apjs, 5, 139

\bibitem[{{Amarsi} {et~al.}(2019){Amarsi}, {Nissen}, {Asplund}, {Lind}, \&
  {Barklem}}]{Amarsi19}
{Amarsi}, A.~M., {Nissen}, P.~E., {Asplund}, M., {Lind}, K., \& {Barklem},
  P.~S. 2019, \aap, 622, L4

\bibitem[{{Amor{\'\i}n} {et~al.}(2016){Amor{\'\i}n}, {Mu{\~n}oz-Tu{\~n}{\'o}n},
  {Aguerri}, \& {Planesas}}]{Amorin16}
{Amor{\'\i}n}, R., {Mu{\~n}oz-Tu{\~n}{\'o}n}, C., {Aguerri}, J.~A.~L., \&
  {Planesas}, P. 2016, \aap, 588, A23

\bibitem[{{Asano} {et~al.}(2013{\natexlab{a}}){Asano}, {Takeuchi}, {Hirashita},
  \& {Inoue}}]{Asano13}
{Asano}, R.~S., {Takeuchi}, T.~T., {Hirashita}, H., \& {Inoue}, A.~K.
  2013{\natexlab{a}}, Earth, Planets and Space, 65, 213

\bibitem[{{Asano} {et~al.}(2013{\natexlab{b}}){Asano}, {Takeuchi}, {Hirashita},
  \& {Nozawa}}]{Asano13b}
{Asano}, R.~S., {Takeuchi}, T.~T., {Hirashita}, H., \& {Nozawa}, T.
  2013{\natexlab{b}}, \mnras, 432, 637

\bibitem[{{Asplund} {et~al.}(2009){Asplund}, {Grevesse}, {Sauval}, \&
  {Scott}}]{Asplund2009}
{Asplund}, M., {Grevesse}, N., {Sauval}, A.~J., \& {Scott}, P. 2009, \araa, 47,
  481

\bibitem[{{Balashev} {et~al.}(2017){Balashev}, {Noterdaeme}, {Rahmani},
  {Klimenko}, {Ledoux}, {Petitjean}, {Srianand}, {Ivanchik}, \&
  {Varshalovich}}]{Balashev17}
{Balashev}, S.~A., {Noterdaeme}, P., {Rahmani}, H., {et~al.} 2017, \mnras, 470,
  2890

\bibitem[{{Bate}(2019)}]{Bate19}
{Bate}, M.~R. 2019, \mnras, 484, 2341

\bibitem[{{Bell} {et~al.}(2006){Bell}, {Roueff}, {Viti}, \&
  {Williams}}]{Bell06}
{Bell}, T.~A., {Roueff}, E., {Viti}, S., \& {Williams}, D.~A. 2006, \mnras,
  371, 1865

\bibitem[{{Bell} {et~al.}(2007){Bell}, {Viti}, \& {Williams}}]{Bell07}
{Bell}, T.~A., {Viti}, S., \& {Williams}, D.~A. 2007, \mnras, 378, 983

\bibitem[{{Berg} {et~al.}(2019){Berg}, {Erb}, {Henry}, {Skillman}, \&
  {McQuinn}}]{Berg19}
{Berg}, D.~A., {Erb}, D.~K., {Henry}, R. B.~C., {Skillman}, E.~D., \&
  {McQuinn}, K. B.~W. 2019, \apj, 874, 93

\bibitem[{{Berg} {et~al.}(2016){Berg}, {Skillman}, {Henry}, {Erb}, \&
  {Carigi}}]{Berg16}
{Berg}, D.~A., {Skillman}, E.~D., {Henry}, R. B.~C., {Erb}, D.~K., \& {Carigi},
  L. 2016, \apj, 827, 126

\bibitem[{{Bialy} {et~al.}(2015){Bialy}, {Sternberg}, {Lee}, {Le Petit}, \&
  {Roueff}}]{Bialy15}
{Bialy}, S., {Sternberg}, A., {Lee}, M.-Y., {Le Petit}, F., \& {Roueff}, E.
  2015, \apj, 809, 122

\bibitem[{{Bigiel} {et~al.}(2020){Bigiel}, {de Looze}, {Krabbe}, {Cormier},
  {Barnes}, {Fischer}, {Bolatto}, {Bryant}, {Colditz}, {Geis}, {Herrera-Camus},
  {Iserlohe}, {Klein}, {Leroy}, {Linz}, {Looney}, {Madden}, {Poglitsch},
  {Stutzki}, \& {Vacca}}]{Bigiel20}
{Bigiel}, F., {de Looze}, I., {Krabbe}, A., {et~al.} 2020, \apj, 903, 30

\bibitem[{{Bisbas} {et~al.}(2012){Bisbas}, {Bell}, {Viti}, {Yates}, \&
  {Barlow}}]{Bisbas12}
{Bisbas}, T.~G., {Bell}, T.~A., {Viti}, S., {Yates}, J., \& {Barlow}, M.~J.
  2012, \mnras, 427, 2100

\bibitem[{{Bisbas} {et~al.}(2015){Bisbas}, {Papadopoulos}, \&
  {Viti}}]{Bisbas15}
{Bisbas}, T.~G., {Papadopoulos}, P.~P., \& {Viti}, S. 2015, \apj, 803, 37

\bibitem[{{Bisbas} {et~al.}(2019){Bisbas}, {Schruba}, \& {van
  Dishoeck}}]{Bisbas19}
{Bisbas}, T.~G., {Schruba}, A., \& {van Dishoeck}, E.~F. 2019, \mnras, 485,
  3097

\bibitem[{{Bisbas} {et~al.}(2021){Bisbas}, {Tan}, \& {Tanaka}}]{Bisbas21}
{Bisbas}, T.~G., {Tan}, J.~C., \& {Tanaka}, K. E.~I. 2021, \mnras, 502, 2701

\bibitem[{{Bisbas} {et~al.}(2017{\natexlab{a}}){Bisbas}, {Tanaka}, {Tan}, {Wu},
  \& {Nakamura}}]{Bisbas17b}
{Bisbas}, T.~G., {Tanaka}, K. E.~I., {Tan}, J.~C., {Wu}, B., \& {Nakamura}, F.
  2017{\natexlab{a}}, \apj, 850, 23

\bibitem[{{Bisbas} {et~al.}(2023){Bisbas}, {van Dishoeck}, {Hu}, \&
  {Schruba}}]{Bisbas23}
{Bisbas}, T.~G., {van Dishoeck}, E.~F., {Hu}, C.-Y., \& {Schruba}, A. 2023,
  \mnras, 519, 729

\bibitem[{{Bisbas} {et~al.}(2017{\natexlab{b}}){Bisbas}, {van Dishoeck},
  {Papadopoulos}, {Sz{\H{u}}cs}, {Bialy}, \& {Zhang}}]{Bisbas17}
{Bisbas}, T.~G., {van Dishoeck}, E.~F., {Papadopoulos}, P.~P., {et~al.}
  2017{\natexlab{b}}, \apj, 839, 90

\bibitem[{{Bisbas} {et~al.}(2024){Bisbas}, {Zhang}, {Gjergo}, {Zhao}, {Luo},
  {Quan}, {Jiang}, {Sun}, {Topkaras}, {Li}, \& {Guo}}]{Bisbas24}
{Bisbas}, T.~G., {Zhang}, Z.-Y., {Gjergo}, E., {et~al.} 2024, \mnras, 527, 8886

\bibitem[{{Bolatto} {et~al.}(2013){Bolatto}, {Wolfire}, \& {Leroy}}]{Bolatto13}
{Bolatto}, A.~D., {Wolfire}, M., \& {Leroy}, A.~K. 2013, \araa, 51, 207

\bibitem[{{Borchert} {et~al.}(2022){Borchert}, {Walch}, {Seifried}, {Clarke},
  {Franeck}, \& {N{\"u}rnberger}}]{Borchert22}
{Borchert}, E.~M.~A., {Walch}, S., {Seifried}, D., {et~al.} 2022, \mnras, 510,
  753

\bibitem[{{Chevance} {et~al.}(2020){Chevance}, {Madden}, {Fischer}, {Vacca},
  {Lebouteiller}, {Fadda}, {Galliano}, {Indebetouw}, {Kruijssen}, {Lee},
  {Poglitsch}, {Polles}, {Cormier}, {Hony}, {Iserlohe}, {Krabbe}, {Meixner},
  {Sabbi}, \& {Zinnecker}}]{Chevance20}
{Chevance}, M., {Madden}, S.~C., {Fischer}, C., {et~al.} 2020, \mnras, 494,
  5279

\bibitem[{{Choban} {et~al.}(2024){Choban}, {Kere{\v{s}}}, {Sandstrom},
  {Hopkins}, {Hayward}, \& {Faucher-Gigu{\`e}re}}]{Choban24}
{Choban}, C.~R., {Kere{\v{s}}}, D., {Sandstrom}, K.~M., {et~al.} 2024, \mnras,
  529, 2356

\bibitem[{{Cooke} {et~al.}(2017){Cooke}, {Pettini}, \& {Steidel}}]{Cooke17}
{Cooke}, R.~J., {Pettini}, M., \& {Steidel}, C.~C. 2017, \mnras, 467, 802

\bibitem[{{Crocker} {et~al.}(2019){Crocker}, {Pellegrini}, {Smith}, {Draine},
  {Wilson}, {Wolfire}, {Armus}, {Brinks}, {Dale}, {Groves}, {Herrera-Camus},
  {Hunt}, {Kennicutt}, {Murphy}, {Sandstrom}, {Schinnerer}, {Rigopoulou},
  {Rosolowsky}, \& {van der Werf}}]{Crocker19}
{Crocker}, A.~F., {Pellegrini}, E., {Smith}, J. D.~T., {et~al.} 2019, \apj,
  887, 105

\bibitem[{{Dabrowski}(1984)}]{Dabrowski84}
{Dabrowski}, I. 1984, Canadian Journal of Physics, 62, 1639

\bibitem[{{Dame} {et~al.}(2001){Dame}, {Hartmann}, \& {Thaddeus}}]{Dame01}
{Dame}, T.~M., {Hartmann}, D., \& {Thaddeus}, P. 2001, \apj, 547, 792

\bibitem[{{Draine}(1978)}]{Draine78}
{Draine}, B.~T. 1978, \apjs, 36, 595

\bibitem[{{Dufour}(1984)}]{Dufour84}
{Dufour}, R.~J. 1984, in IAU Symposium, Vol. 108, Structure and Evolution of
  the Magellanic Clouds, ed. S.~{van den Bergh} \& K.~S.~D. {de Boer}, 353--361

\bibitem[{{Dunne} {et~al.}(2022){Dunne}, {Maddox}, {Papadopoulos}, {Ivison}, \&
  {Gomez}}]{Dunne22}
{Dunne}, L., {Maddox}, S.~J., {Papadopoulos}, P.~P., {Ivison}, R.~J., \&
  {Gomez}, H.~L. 2022, \mnras, 517, 962

\bibitem[{{Dunne} {et~al.}(2021){Dunne}, {Maddox}, {Vlahakis}, \&
  {Gomez}}]{Dunne21}
{Dunne}, L., {Maddox}, S.~J., {Vlahakis}, C., \& {Gomez}, H.~L. 2021, \mnras,
  501, 2573

\bibitem[{{Elmegreen} {et~al.}(2013){Elmegreen}, {Rubio}, {Hunter}, {Verdugo},
  {Brinks}, \& {Schruba}}]{Elmegreen13}
{Elmegreen}, B.~G., {Rubio}, M., {Hunter}, D.~A., {et~al.} 2013, \nat, 495, 487

\bibitem[{{Esteban} {et~al.}(2009){Esteban}, {Bresolin}, {Peimbert},
  {Garc{\'\i}a-Rojas}, {Peimbert}, \& {Mesa-Delgado}}]{Esteban09}
{Esteban}, C., {Bresolin}, F., {Peimbert}, M., {et~al.} 2009, \apj, 700, 654

\bibitem[{{Esteban} {et~al.}(2014){Esteban}, {Garc{\'\i}a-Rojas}, {Carigi},
  {Peimbert}, {Bresolin}, {L{\'o}pez-S{\'a}nchez}, \&
  {Mesa-Delgado}}]{Esteban14}
{Esteban}, C., {Garc{\'\i}a-Rojas}, J., {Carigi}, L., {et~al.} 2014, \mnras,
  443, 624

\bibitem[{{Esteban} {et~al.}(2002){Esteban}, {Peimbert}, {Torres-Peimbert}, \&
  {Rodr{\'\i}guez}}]{Esteban02}
{Esteban}, C., {Peimbert}, M., {Torres-Peimbert}, S., \& {Rodr{\'\i}guez}, M.
  2002, \apj, 581, 241

\bibitem[{{Ferland} {et~al.}(2017){Ferland}, {Chatzikos}, {Guzm{\'a}n},
  {Lykins}, {van Hoof}, {Williams}, {Abel}, {Badnell}, {Keenan}, {Porter}, \&
  {Stancil}}]{Ferland17}
{Ferland}, G.~J., {Chatzikos}, M., {Guzm{\'a}n}, F., {et~al.} 2017, \rmxaa, 53,
  385

\bibitem[{{Froebrich} \& {Rowles}(2010)}]{Froebrich10}
{Froebrich}, D. \& {Rowles}, J. 2010, \mnras, 406, 1350

\bibitem[{{Gaches} {et~al.}(2019){Gaches}, {Offner}, \& {Bisbas}}]{Gaches19b}
{Gaches}, B. A.~L., {Offner}, S. S.~R., \& {Bisbas}, T.~G. 2019, \apj, 883, 190

\bibitem[{{Galametz} {et~al.}(2011){Galametz}, {Madden}, {Galliano}, {Hony},
  {Bendo}, \& {Sauvage}}]{Galametz11}
{Galametz}, M., {Madden}, S.~C., {Galliano}, F., {et~al.} 2011, \aap, 532, A56

\bibitem[{{Galliano} {et~al.}(2021){Galliano}, {Nersesian}, {Bianchi}, {De
  Looze}, {Roychowdhury}, {Baes}, {Casasola}, {Cassar{\'a}}, {Dobbels},
  {Fritz}, {Galametz}, {Jones}, {Madden}, {Mosenkov}, {Xilouris}, \&
  {Ysard}}]{Galliano21}
{Galliano}, F., {Nersesian}, A., {Bianchi}, S., {et~al.} 2021, \aap, 649, A18

\bibitem[{{Garc{\'\i}a-Rojas} \& {Esteban}(2007)}]{Garcia-Rojas07}
{Garc{\'\i}a-Rojas}, J. \& {Esteban}, C. 2007, \apj, 670, 457

\bibitem[{{Garnett} {et~al.}(1995){Garnett}, {Skillman}, {Dufour}, {Peimbert},
  {Torres-Peimbert}, {Terlevich}, {Terlevich}, \& {Shields}}]{Garnett95}
{Garnett}, D.~R., {Skillman}, E.~D., {Dufour}, R.~J., {et~al.} 1995, \apj, 443,
  64

\bibitem[{{Genzel} {et~al.}(2012){Genzel}, {Tacconi}, {Combes}, {Bolatto},
  {Neri}, {Sternberg}, {Cooper}, {Bouch{\'e}}, {Bournaud}, {Burkert},
  {Comerford}, {Cox}, {Davis}, {F{\"o}rster Schreiber}, {Garcia-Burillo},
  {Gracia-Carpio}, {Lutz}, {Naab}, {Newman}, {Saintonge}, {Shapiro}, {Shapley},
  \& {Weiner}}]{Genzel12}
{Genzel}, R., {Tacconi}, L.~J., {Combes}, F., {et~al.} 2012, \apj, 746, 69

\bibitem[{{Giveon} {et~al.}(2002){Giveon}, {Sternberg}, {Lutz}, {Feuchtgruber},
  \& {Pauldrach}}]{Giveon02}
{Giveon}, U., {Sternberg}, A., {Lutz}, D., {Feuchtgruber}, H., \& {Pauldrach},
  A.~W.~A. 2002, \apj, 566, 880

\bibitem[{{Glover} \& {Clark}(2016)}]{Glover16}
{Glover}, S. C.~O. \& {Clark}, P.~C. 2016, \mnras, 456, 3596

\bibitem[{{Glover} {et~al.}(2015){Glover}, {Clark}, {Micic}, \&
  {Molina}}]{Glover15}
{Glover}, S. C.~O., {Clark}, P.~C., {Micic}, M., \& {Molina}, F. 2015, \mnras,
  448, 1607

\bibitem[{{Glover} {et~al.}(2010){Glover}, {Federrath}, {Mac Low}, \&
  {Klessen}}]{Glover10}
{Glover}, S.~C.~O., {Federrath}, C., {Mac Low}, M.~M., \& {Klessen}, R.~S.
  2010, \mnras, 404, 2

\bibitem[{{Gong} {et~al.}(2018){Gong}, {Ostriker}, \& {Kim}}]{Gong18}
{Gong}, M., {Ostriker}, E.~C., \& {Kim}, C.-G. 2018, \apj, 858, 16

\bibitem[{{Gong} {et~al.}(2020){Gong}, {Ostriker}, {Kim}, \& {Kim}}]{Gong20}
{Gong}, M., {Ostriker}, E.~C., {Kim}, C.-G., \& {Kim}, J.-G. 2020, \apj, 903,
  142

\bibitem[{{Goodman} {et~al.}(2009){Goodman}, {Pineda}, \& {Schnee}}]{Goodman09}
{Goodman}, A.~A., {Pineda}, J.~E., \& {Schnee}, S.~L. 2009, \apj, 692, 91

\bibitem[{{Harrington} {et~al.}(2021){Harrington}, {Weiss}, {Yun}, {Magnelli},
  {Sharon}, {Leung}, {Vishwas}, {Wang}, {Frayer}, {Jim{\'e}nez-Andrade}, {Liu},
  {Garc{\'\i}a}, {Romano-D{\'\i}az}, {Frye}, {Jarugula}, {B{\u{a}}descu},
  {Berman}, {Dannerbauer}, {D{\'\i}az-S{\'a}nchez}, {Grassitelli},
  {Kamieneski}, {Kim}, {Kirkpatrick}, {Lowenthal}, {Messias}, {Puschnig},
  {Stacey}, {Torne}, \& {Bertoldi}}]{Harrington21}
{Harrington}, K.~C., {Weiss}, A., {Yun}, M.~S., {et~al.} 2021, \apj, 908, 95

\bibitem[{{Heintz} \& {Watson}(2020)}]{Heintz20}
{Heintz}, K.~E. \& {Watson}, D. 2020, \apjl, 889, L7

\bibitem[{{Herrera-Camus} {et~al.}(2012){Herrera-Camus}, {Fisher}, {Bolatto},
  {Leroy}, {Walter}, {Gordon}, {Roman-Duval}, {Donaldson}, {Mel{\'e}ndez}, \&
  {Cannon}}]{Herrera12}
{Herrera-Camus}, R., {Fisher}, D.~B., {Bolatto}, A.~D., {et~al.} 2012, \apj,
  752, 112

\bibitem[{{Heyer} \& {Dame}(2015)}]{Heyer15}
{Heyer}, M. \& {Dame}, T.~M. 2015, \araa, 53, 583

\bibitem[{{Hollenbach} \& {Tielens}(1999)}]{Hollenbach99}
{Hollenbach}, D.~J. \& {Tielens}, A.~G.~G.~M. 1999, Reviews of Modern Physics,
  71, 173

\bibitem[{{Hu} {et~al.}(2022){Hu}, {Schruba}, {Sternberg}, \& {van
  Dishoeck}}]{Hu22}
{Hu}, C.-Y., {Schruba}, A., {Sternberg}, A., \& {van Dishoeck}, E.~F. 2022,
  \apj, 931, 28

\bibitem[{{Hu} {et~al.}(2021){Hu}, {Sternberg}, \& {van Dishoeck}}]{Hu21}
{Hu}, C.-Y., {Sternberg}, A., \& {van Dishoeck}, E.~F. 2021, \apj, 920, 44

\bibitem[{{Hunt} {et~al.}(2023){Hunt}, {Belfiore}, {Lelli}, {Draine},
  {Marasco}, {Garc{\'\i}a-Burillo}, {Venturi}, {Combes}, {Wei{\ss}}, {Henkel},
  {Menten}, {Annibali}, {Casasola}, {Cignoni}, {McLeod}, {Tosi}, {Beltr{\'a}n},
  {Concas}, {Cresci}, {Ginolfi}, {Kumari}, \& {Mannucci}}]{Hunt23}
{Hunt}, L.~K., {Belfiore}, F., {Lelli}, F., {et~al.} 2023, \aap, 675, A64

\bibitem[{{Hunt} {et~al.}(2015){Hunt}, {Garc{\'\i}a-Burillo}, {Casasola},
  {Caselli}, {Combes}, {Henkel}, {Lundgren}, {Maiolino}, {Menten}, {Testi}, \&
  {Weiss}}]{Hunt15}
{Hunt}, L.~K., {Garc{\'\i}a-Burillo}, S., {Casasola}, V., {et~al.} 2015, \aap,
  583, A114

\bibitem[{{Hunter} {et~al.}(2024){Hunter}, {Elmegreen}, \& {Madden}}]{Hunter24}
{Hunter}, D.~A., {Elmegreen}, B.~G., \& {Madden}, S.~C. 2024, \araa, 62, 113

\bibitem[{{Israel}(1997)}]{Israel97}
{Israel}, F.~P. 1997, \aap, 328, 471

\bibitem[{{Israel} \& {Maloney}(2011)}]{Israel11}
{Israel}, F.~P. \& {Maloney}, P.~R. 2011, \aap, 531, A19

\bibitem[{{Izotov} {et~al.}(2023){Izotov}, {Schaerer}, {Worseck}, {Berg},
  {Chisholm}, {Ravindranath}, \& {Thuan}}]{Izotov23}
{Izotov}, Y.~I., {Schaerer}, D., {Worseck}, G., {et~al.} 2023, \mnras, 522,
  1228

\bibitem[{{Jiao} {et~al.}(2021){Jiao}, {Gao}, \& {Zhao}}]{Jiao21}
{Jiao}, Q., {Gao}, Y., \& {Zhao}, Y. 2021, \mnras, 504, 2360

\bibitem[{{Jiao} {et~al.}(2019){Jiao}, {Zhao}, {Lu}, {Gao}, {Salak}, {Zhu},
  {Zhang}, {Jiang}, \& {Tan}}]{Jiao19}
{Jiao}, Q., {Zhao}, Y., {Lu}, N., {et~al.} 2019, \apj, 880, 133

\bibitem[{{Jiao} {et~al.}(2017){Jiao}, {Zhao}, {Zhu}, {Lu}, {Gao}, \&
  {Zhang}}]{Jiao17}
{Jiao}, Q., {Zhao}, Y., {Zhu}, M., {et~al.} 2017, \apjl, 840, L18

\bibitem[{{Kainulainen} {et~al.}(2009){Kainulainen}, {Beuther}, {Henning}, \&
  {Plume}}]{Kainulainen09}
{Kainulainen}, J., {Beuther}, H., {Henning}, T., \& {Plume}, R. 2009, \aap,
  508, L35

\bibitem[{{Kennicutt}(1998)}]{Kennicutt98}
{Kennicutt}, Jr., R.~C. 1998, \apj, 498, 541

\bibitem[{{Komugi} {et~al.}(2023){Komugi}, {Inaba}, \& {Shindou}}]{Komugi23}
{Komugi}, S., {Inaba}, M., \& {Shindou}, T. 2023, \pasj, 75, 1337

\bibitem[{{Konstantopoulou} {et~al.}(2024){Konstantopoulou}, {De Cia},
  {Ledoux}, {Krogager}, {Mattsson}, {Watson}, {Heintz}, {P{\'e}roux},
  {Noterdaeme}, {Andersen}, {Fynbo}, {Jermann}, \&
  {Ramburuth-Hurt}}]{Konstantopoulou24}
{Konstantopoulou}, C., {De Cia}, A., {Ledoux}, C., {et~al.} 2024, \aap, 681,
  A64

\bibitem[{{Lada} \& {Blitz}(1988)}]{Lada88}
{Lada}, E.~A. \& {Blitz}, L. 1988, \apjl, 326, L69

\bibitem[{{Langer} {et~al.}(2014){Langer}, {Velusamy}, {Pineda}, {Willacy}, \&
  {Goldsmith}}]{Langer14}
{Langer}, W.~D., {Velusamy}, T., {Pineda}, J.~L., {Willacy}, K., \&
  {Goldsmith}, P.~F. 2014, \aap, 561, A122

\bibitem[{{Lelli} {et~al.}(2023){Lelli}, {Zhang}, {Bisbas}, {Lin},
  {Papadopoulos}, {Schombert}, {Di Teodoro}, {Marasco}, \& {McGaugh}}]{Lelli23}
{Lelli}, F., {Zhang}, Z.-Y., {Bisbas}, T.~G., {et~al.} 2023, \aap, 672, A106

\bibitem[{{Leroy} {et~al.}(2011){Leroy}, {Bolatto}, {Gordon}, {Sandstrom},
  {Gratier}, {Rosolowsky}, {Engelbracht}, {Mizuno}, {Corbelli}, {Fukui}, \&
  {Kawamura}}]{Leroy11}
{Leroy}, A.~K., {Bolatto}, A., {Gordon}, K., {et~al.} 2011, \apj, 737, 12

\bibitem[{{Leroy} {et~al.}(2016){Leroy}, {Hughes}, {Schruba}, {Rosolowsky},
  {Blanc}, {Bolatto}, {Colombo}, {Escala}, {Kramer}, {Kruijssen}, {Meidt},
  {Pety}, {Querejeta}, {Sandstrom}, {Schinnerer}, {Sliwa}, \&
  {Usero}}]{Leroy16}
{Leroy}, A.~K., {Hughes}, A., {Schruba}, A., {et~al.} 2016, \apj, 831, 16

\bibitem[{{Lo} {et~al.}(2014){Lo}, {Cunningham}, {Jones}, {Bronfman}, {Cortes},
  {Simon}, {Lowe}, {Fissel}, \& {Novak}}]{Lo14}
{Lo}, N., {Cunningham}, M.~R., {Jones}, P.~A., {et~al.} 2014, \apjl, 797, L17

\bibitem[{{L{\'o}pez-S{\'a}nchez} {et~al.}(2007){L{\'o}pez-S{\'a}nchez},
  {Esteban}, {Garc{\'\i}a-Rojas}, {Peimbert}, \&
  {Rodr{\'\i}guez}}]{Lopez-Sanchez07}
{L{\'o}pez-S{\'a}nchez}, {\'A}.~R., {Esteban}, C., {Garc{\'\i}a-Rojas}, J.,
  {Peimbert}, M., \& {Rodr{\'\i}guez}, M. 2007, \apj, 656, 168

\bibitem[{{Luo} {et~al.}(2020){Luo}, {Li}, {Tang}, {Dawson}, {Dickey},
  {Bronfman}, {Qin}, {Gibson}, {Plambeck}, {Finger}, {Green}, {Mardones},
  {Koo}, \& {Lo}}]{Luo20}
{Luo}, G., {Li}, D., {Tang}, N., {et~al.} 2020, \apjl, 889, L4

\bibitem[{{Luo} {et~al.}(2024){Luo}, {Li}, {Zhang}, {Bisbas}, {Tang}, {Lin},
  {Sun}, {Zuo}, \& {Zhou}}]{Luo24}
{Luo}, G., {Li}, D., {Zhang}, Z.-Y., {et~al.} 2024, \aap, 685, L12

\bibitem[{{Ma} {et~al.}(2022){Ma}, {Wang}, {Zhang}, {Wang}, {Zhang}, {Liu},
  {Li}, {Zheng}, {Yuan}, \& {Yang}}]{Ma22}
{Ma}, Y., {Wang}, H., {Zhang}, M., {et~al.} 2022, \apjs, 262, 16

\bibitem[{{Madden} {et~al.}(2020){Madden}, {Cormier}, {Hony}, {Lebouteiller},
  {Abel}, {Galametz}, {De Looze}, {Chevance}, {Polles}, {Lee}, {Galliano},
  {Lambert-Huyghe}, {Hu}, \& {Ramambason}}]{Madden20}
{Madden}, S.~C., {Cormier}, D., {Hony}, S., {et~al.} 2020, \aap, 643, A141

\bibitem[{{Madden} {et~al.}(2013){Madden}, {R{\'e}my-Ruyer}, {Galametz},
  {Cormier}, {Lebouteiller}, {Galliano}, {Hony}, {Bendo}, {Smith}, {Pohlen},
  {Roussel}, {Sauvage}, {Wu}, {Sturm}, {Poglitsch}, {Contursi}, {Doublier},
  {Baes}, {Barlow}, {Boselli}, {Boquien}, {Carlson}, {Ciesla}, {Cooray},
  {Cortese}, {de Looze}, {Irwin}, {Isaak}, {Kamenetzky}, {Karczewski}, {Lu},
  {MacHattie}, {O'Halloran}, {Parkin}, {Rangwala}, {Schirm}, {Schulz},
  {Spinoglio}, {Vaccari}, {Wilson}, \& {Wozniak}}]{Madden13}
{Madden}, S.~C., {R{\'e}my-Ruyer}, A., {Galametz}, M., {et~al.} 2013, \pasp,
  125, 600

\bibitem[{{Magdis} {et~al.}(2012){Magdis}, {Daddi}, {B{\'e}thermin}, {Sargent},
  {Elbaz}, {Pannella}, {Dickinson}, {Dannerbauer}, {da Cunha}, {Walter},
  {Rigopoulou}, {Charmandaris}, {Hwang}, \& {Kartaltepe}}]{Magdis12}
{Magdis}, G.~E., {Daddi}, E., {B{\'e}thermin}, M., {et~al.} 2012, \apj, 760, 6

\bibitem[{{Magdis} {et~al.}(2011){Magdis}, {Daddi}, {Elbaz}, {Sargent},
  {Dickinson}, {Dannerbauer}, {Aussel}, {Walter}, {Hwang}, {Charmandaris},
  {Hodge}, {Riechers}, {Rigopoulou}, {Carilli}, {Pannella}, {Mullaney},
  {Leiton}, \& {Scott}}]{Magdis11}
{Magdis}, G.~E., {Daddi}, E., {Elbaz}, D., {et~al.} 2011, \apjl, 740, L15

\bibitem[{{Maiolino} \& {Mannucci}(2019)}]{Maiolino19}
{Maiolino}, R. \& {Mannucci}, F. 2019, \aapr, 27, 3

\bibitem[{{Marks} {et~al.}(2012){Marks}, {Kroupa}, {Dabringhausen}, \&
  {Pawlowski}}]{Marks12}
{Marks}, M., {Kroupa}, P., {Dabringhausen}, J., \& {Pawlowski}, M.~S. 2012,
  \mnras, 422, 2246

\bibitem[{{McElroy} {et~al.}(2013){McElroy}, {Walsh}, {Markwick}, {Cordiner},
  {Smith}, \& {Millar}}]{McElroy13}
{McElroy}, D., {Walsh}, C., {Markwick}, A.~J., {et~al.} 2013, \aap, 550, A36

\bibitem[{{McKee} \& {Ostriker}(2007)}]{McKee07}
{McKee}, C.~F. \& {Ostriker}, E.~C. 2007, \araa, 45, 565

\bibitem[{{Michiyama} {et~al.}(2021){Michiyama}, {Saito}, {Tadaki}, {Ueda},
  {Zhuang}, {Molina}, {Lee}, {Wang}, {Bolatto}, {Iono}, {Nakanishi}, {Izumi},
  {Yamashita}, \& {Ho}}]{Michiyama21}
{Michiyama}, T., {Saito}, T., {Tadaki}, K.-i., {et~al.} 2021, \apjs, 257, 28

\bibitem[{{Michiyama} {et~al.}(2020){Michiyama}, {Ueda}, {Tadaki}, {Bolatto},
  {Molina}, {Saito}, {Yamashita}, {Zhuang}, {Nakanishi}, {Iono}, {Wang}, \&
  {Ho}}]{Michiyama20}
{Michiyama}, T., {Ueda}, J., {Tadaki}, K.-i., {et~al.} 2020, \apjl, 897, L19

\bibitem[{{Nicholls} {et~al.}(2017){Nicholls}, {Sutherland}, {Dopita},
  {Kewley}, \& {Groves}}]{Nicholls2017}
{Nicholls}, D.~C., {Sutherland}, R.~S., {Dopita}, M.~A., {Kewley}, L.~J., \&
  {Groves}, B.~A. 2017, \mnras, 466, 4403

\bibitem[{{Nieten} {et~al.}(2006){Nieten}, {Neininger}, {Gu{\'e}lin},
  {Ungerechts}, {Lucas}, {Berkhuijsen}, {Beck}, \& {Wielebinski}}]{Nieten06}
{Nieten}, C., {Neininger}, N., {Gu{\'e}lin}, M., {et~al.} 2006, \aap, 453, 459

\bibitem[{{Offner} {et~al.}(2014){Offner}, {Bisbas}, {Bell}, \&
  {Viti}}]{Offner14}
{Offner}, S.~S.~R., {Bisbas}, T.~G., {Bell}, T.~A., \& {Viti}, S. 2014, \mnras,
  440, L81

\bibitem[{{Pagel}(2009)}]{Pagel09}
{Pagel}, B. E.~J. 2009, {Nucleosynthesis and Chemical Evolution of Galaxies}

\bibitem[{{Papadopoulos} {et~al.}(2018){Papadopoulos}, {Bisbas}, \&
  {Zhang}}]{Papadopoulos18}
{Papadopoulos}, P.~P., {Bisbas}, T.~G., \& {Zhang}, Z.-Y. 2018, \mnras, 478,
  1716

\bibitem[{{Papadopoulos} {et~al.}(2004){Papadopoulos}, {Thi}, \&
  {Viti}}]{Papadopoulos04}
{Papadopoulos}, P.~P., {Thi}, W.~F., \& {Viti}, S. 2004, \mnras, 351, 147

\bibitem[{{Pilyugin} \& {Thuan}(2005)}]{Pilyugin05}
{Pilyugin}, L.~S. \& {Thuan}, T.~X. 2005, \apj, 631, 231

\bibitem[{{Popping} \& {P{\'e}roux}(2022)}]{Popping22}
{Popping}, G. \& {P{\'e}roux}, C. 2022, \mnras, 513, 1531

\bibitem[{{Pradhan} {et~al.}(2011){Pradhan}, {Murthy}, \& {Pathak}}]{Pradhan11}
{Pradhan}, A.~C., {Murthy}, J., \& {Pathak}, A. 2011, \apj, 743, 80

\bibitem[{{Ramambason} {et~al.}(2024){Ramambason}, {Lebouteiller}, {Madden},
  {Galliano}, {Richardson}, {Saintonge}, {De Looze}, {Chevance}, {Abel},
  {Hernandez}, \& {Braine}}]{Ramambason24}
{Ramambason}, L., {Lebouteiller}, V., {Madden}, S.~C., {et~al.} 2024, \aap,
  681, A14

\bibitem[{{Ravindranath} {et~al.}(2020){Ravindranath}, {Monroe}, {Jaskot},
  {Ferguson}, \& {Tumlinson}}]{Ravindranath20}
{Ravindranath}, S., {Monroe}, T., {Jaskot}, A., {Ferguson}, H.~C., \&
  {Tumlinson}, J. 2020, \apj, 896, 170

\bibitem[{{R{\'e}my-Ruyer} {et~al.}(2014){R{\'e}my-Ruyer}, {Madden},
  {Galliano}, {Galametz}, {Takeuchi}, {Asano}, {Zhukovska}, {Lebouteiller},
  {Cormier}, {Jones}, {Bocchio}, {Baes}, {Bendo}, {Boquien}, {Boselli},
  {DeLooze}, {Doublier-Pritchard}, {Hughes}, {Karczewski}, \&
  {Spinoglio}}]{RemyRuyer14}
{R{\'e}my-Ruyer}, A., {Madden}, S.~C., {Galliano}, F., {et~al.} 2014, \aap,
  563, A31

\bibitem[{{R{\'e}my-Ruyer} {et~al.}(2013){R{\'e}my-Ruyer}, {Madden},
  {Galliano}, {Hony}, {Sauvage}, {Bendo}, {Roussel}, {Pohlen}, {Smith},
  {Galametz}, {Cormier}, {Lebouteiller}, {Wu}, {Baes}, {Barlow}, {Boquien},
  {Boselli}, {Ciesla}, {De Looze}, {Karczewski}, {Panuzzo}, {Spinoglio},
  {Vaccari}, \& {Wilson}}]{RemyRuyer13}
{R{\'e}my-Ruyer}, A., {Madden}, S.~C., {Galliano}, F., {et~al.} 2013, \aap,
  557, A95

\bibitem[{{Rice} {et~al.}(2016){Rice}, {Goodman}, {Bergin}, {Beaumont}, \&
  {Dame}}]{Rice16}
{Rice}, T.~S., {Goodman}, A.~A., {Bergin}, E.~A., {Beaumont}, C., \& {Dame},
  T.~M. 2016, \apj, 822, 52

\bibitem[{{R{\"o}llig} {et~al.}(2007){R{\"o}llig}, {Abel}, {Bell}, {Bensch},
  {Black}, {Ferland}, {Jonkheid}, {Kamp}, {Kaufman}, {Le Bourlot}, {Le Petit},
  {Meijerink}, {Morata}, {Ossenkopf}, {Roueff}, {Shaw}, {Spaans}, {Sternberg},
  {Stutzki}, {Thi}, {van Dishoeck}, {van Hoof}, {Viti}, \&
  {Wolfire}}]{Roellig07}
{R{\"o}llig}, M., {Abel}, N.~P., {Bell}, T., {et~al.} 2007, \aap, 467, 187

\bibitem[{{Romano}(2022)}]{Romano22}
{Romano}, D. 2022, \aapr, 30, 7

\bibitem[{{Safranek-Shrader} {et~al.}(2017){Safranek-Shrader}, {Krumholz},
  {Kim}, {Ostriker}, {Klein}, {Li}, {McKee}, \& {Stone}}]{Safranek17}
{Safranek-Shrader}, C., {Krumholz}, M.~R., {Kim}, C.-G., {et~al.} 2017, \mnras,
  465, 885

\bibitem[{{Sandstrom} {et~al.}(2013){Sandstrom}, {Leroy}, {Walter}, {Bolatto},
  {Croxall}, {Draine}, {Wilson}, {Wolfire}, {Calzetti}, {Kennicutt}, {Aniano},
  {Donovan Meyer}, {Usero}, {Bigiel}, {Brinks}, {de Blok}, {Crocker}, {Dale},
  {Engelbracht}, {Galametz}, {Groves}, {Hunt}, {Koda}, {Kreckel}, {Linz},
  {Meidt}, {Pellegrini}, {Rix}, {Roussel}, {Schinnerer}, {Schruba}, {Schuster},
  {Skibba}, {van der Laan}, {Appleton}, {Armus}, {Brandl}, {Gordon}, {Hinz},
  {Krause}, {Montiel}, {Sauvage}, {Schmiedeke}, {Smith}, \&
  {Vigroux}}]{Sandstrom13}
{Sandstrom}, K.~M., {Leroy}, A.~K., {Walter}, F., {et~al.} 2013, \apj, 777, 5

\bibitem[{{Savitzky} \& {Golay}(1964)}]{Savitzky64}
{Savitzky}, A. \& {Golay}, M.~J.~E. 1964, Analytical Chemistry, 36, 1627

\bibitem[{{Schneider} {et~al.}(2016){Schneider}, {Bontemps}, {Motte},
  {Ossenkopf}, {Klessen}, {Simon}, {Fechtenbaum}, {Herpin}, {Tremblin},
  {Csengeri}, {Myers}, {Hill}, {Cunningham}, \& {Federrath}}]{Schneider16}
{Schneider}, N., {Bontemps}, S., {Motte}, F., {et~al.} 2016, \aap, 587, A74

\bibitem[{{Schruba} {et~al.}(2017){Schruba}, {Leroy}, {Kruijssen}, {Bigiel},
  {Bolatto}, {de Blok}, {Tacconi}, {van Dishoeck}, \& {Walter}}]{Schruba17}
{Schruba}, A., {Leroy}, A.~K., {Kruijssen}, J.~M.~D., {et~al.} 2017, \apj, 835,
  278

\bibitem[{{Schruba} {et~al.}(2012){Schruba}, {Leroy}, {Walter}, {Bigiel},
  {Brinks}, {de Blok}, {Kramer}, {Rosolowsky}, {Sandstrom}, {Schuster},
  {Usero}, {Weiss}, \& {Wiesemeyer}}]{Schruba12}
{Schruba}, A., {Leroy}, A.~K., {Walter}, F., {et~al.} 2012, \aj, 143, 138

\bibitem[{{Scoville} {et~al.}(2016){Scoville}, {Sheth}, {Aussel}, {Vanden
  Bout}, {Capak}, {Bongiorno}, {Casey}, {Murchikova}, {Koda},
  {{\'A}lvarez-M{\'a}rquez}, {Lee}, {Laigle}, {McCracken}, {Ilbert}, {Pope},
  {Sanders}, {Chu}, {Toft}, {Ivison}, \& {Manohar}}]{Scoville16}
{Scoville}, N., {Sheth}, K., {Aussel}, H., {et~al.} 2016, \apj, 820, 83

\bibitem[{{Seifried} {et~al.}(2020){Seifried}, {Haid}, {Walch}, {Borchert}, \&
  {Bisbas}}]{Seifried20}
{Seifried}, D., {Haid}, S., {Walch}, S., {Borchert}, E.~M.~A., \& {Bisbas},
  T.~G. 2020, \mnras, 492, 1465

\bibitem[{{Seifried} {et~al.}(2017){Seifried}, {Walch}, {Girichidis}, {Naab},
  {W{\"u}nsch}, {Klessen}, {Glover}, {Peters}, \& {Clark}}]{Seifried17}
{Seifried}, D., {Walch}, S., {Girichidis}, P., {et~al.} 2017, \mnras, 472, 4797

\bibitem[{{Shapley} {et~al.}(2020){Shapley}, {Cullen}, {Dunlop}, {McLure},
  {Kriek}, {Reddy}, \& {Sanders}}]{Shapley20}
{Shapley}, A.~E., {Cullen}, F., {Dunlop}, J.~S., {et~al.} 2020, \apjl, 903, L16

\bibitem[{{Sharda} {et~al.}(2023{\natexlab{a}}){Sharda}, {Amarsi}, {Grasha},
  {Krumholz}, {Yong}, {Chiaki}, {Roy}, \& {Nordlander}}]{Sharda23cor}
{Sharda}, P., {Amarsi}, A.~M., {Grasha}, K., {et~al.} 2023{\natexlab{a}},
  \mnras, 525, 3316

\bibitem[{{Sharda} {et~al.}(2023{\natexlab{b}}){Sharda}, {Amarsi}, {Grasha},
  {Krumholz}, {Yong}, {Chiaki}, {Roy}, \& {Nordlander}}]{Sharda23}
{Sharda}, P., {Amarsi}, A.~M., {Grasha}, K., {et~al.} 2023{\natexlab{b}},
  \mnras, 518, 3985

\bibitem[{{Sharda} \& {Krumholz}(2022)}]{Sharda22}
{Sharda}, P. \& {Krumholz}, M.~R. 2022, \mnras, 509, 1959

\bibitem[{{Shi} {et~al.}(2015){Shi}, {Wang}, {Zhang}, {Gao}, {Armus}, {Helou},
  {Gu}, \& {Stierwalt}}]{Shi15}
{Shi}, Y., {Wang}, J., {Zhang}, Z.-Y., {et~al.} 2015, \apjl, 804, L11

\bibitem[{{Shi} {et~al.}(2016){Shi}, {Wang}, {Zhang}, {Gao}, {Hao}, {Xia}, \&
  {Gu}}]{Shi16}
{Shi}, Y., {Wang}, J., {Zhang}, Z.-Y., {et~al.} 2016, Nature Communications, 7,
  13789

\bibitem[{{Skalidis} {et~al.}(2024){Skalidis}, {Goldsmith}, {Hopkins}, \&
  {Ponnada}}]{Skalidis24}
{Skalidis}, R., {Goldsmith}, P.~F., {Hopkins}, P.~F., \& {Ponnada}, S.~B. 2024,
  \aap, 682, A161

\bibitem[{{Smith} {et~al.}(2014){Smith}, {Glover}, {Clark}, {Klessen}, \&
  {Springel}}]{Smith14}
{Smith}, R.~J., {Glover}, S. C.~O., {Clark}, P.~C., {Klessen}, R.~S., \&
  {Springel}, V. 2014, \mnras, 441, 1628

\bibitem[{{Sofia} {et~al.}(2004){Sofia}, {Lauroesch}, {Meyer}, \&
  {Cartledge}}]{Sofia04}
{Sofia}, U.~J., {Lauroesch}, J.~T., {Meyer}, D.~M., \& {Cartledge}, S. I.~B.
  2004, \apj, 605, 272

\bibitem[{{Spilker} {et~al.}(2021){Spilker}, {Kainulainen}, \&
  {Orkisz}}]{Spilker21}
{Spilker}, A., {Kainulainen}, J., \& {Orkisz}, J. 2021, \aap, 653, A63

\bibitem[{{Sternberg} {et~al.}(2014){Sternberg}, {Le Petit}, {Roueff}, \& {Le
  Bourlot}}]{Sternberg14}
{Sternberg}, A., {Le Petit}, F., {Roueff}, E., \& {Le Bourlot}, J. 2014, \apj,
  790, 10

\bibitem[{{Strong} \& {Mattox}(1996)}]{Strong96}
{Strong}, A.~W. \& {Mattox}, J.~R. 1996, \aap, 308, L21

\bibitem[{{Sz{\H{u}}cs} {et~al.}(2016){Sz{\H{u}}cs}, {Glover}, \&
  {Klessen}}]{Szucs16}
{Sz{\H{u}}cs}, L., {Glover}, S. C.~O., \& {Klessen}, R.~S. 2016, \mnras, 460,
  82

\bibitem[{{Tacconi} {et~al.}(2018){Tacconi}, {Genzel}, {Saintonge}, {Combes},
  {Garc{\'\i}a-Burillo}, {Neri}, {Bolatto}, {Contini}, {F{\"o}rster Schreiber},
  {Lilly}, {Lutz}, {Wuyts}, {Accurso}, {Boissier}, {Boone}, {Bouch{\'e}},
  {Bournaud}, {Burkert}, {Carollo}, {Cooper}, {Cox}, {Feruglio}, {Freundlich},
  {Herrera-Camus}, {Juneau}, {Lippa}, {Naab}, {Renzini}, {Salome}, {Sternberg},
  {Tadaki}, {{\"U}bler}, {Walter}, {Weiner}, \& {Weiss}}]{Tacconi18}
{Tacconi}, L.~J., {Genzel}, R., {Saintonge}, A., {et~al.} 2018, \apj, 853, 179

\bibitem[{{Tanvir} \& {Krumholz}(2024)}]{Tanvir24}
{Tanvir}, T.~S. \& {Krumholz}, M.~R. 2024, \mnras, 527, 7306

\bibitem[{{Tielens}(2005)}]{Tielens05}
{Tielens}, A.~G.~G.~M. 2005, {The Physics and Chemistry of the Interstellar
  Medium}

\bibitem[{{Tielens} \& {Hollenbach}(1985)}]{Tielens85}
{Tielens}, A.~G.~G.~M. \& {Hollenbach}, D. 1985, \apj, 291, 722

\bibitem[{{Tokuda} {et~al.}(2021){Tokuda}, {Kondo}, {Ohno}, {Konishi}, {Sano},
  {Tsuge}, {Zahorecz}, {Goto}, {Neelamkodan}, {Wong}, {Sewi{\l}o}, {Fukushima},
  {Takekoshi}, {Muraoka}, {Kawamura}, {Tachihara}, {Fukui}, \&
  {Onishi}}]{Tokuda21}
{Tokuda}, K., {Kondo}, H., {Ohno}, T., {et~al.} 2021, \apj, 922, 171

\bibitem[{{Trainor} {et~al.}(2016){Trainor}, {Strom}, {Steidel}, \&
  {Rudie}}]{Trainor16}
{Trainor}, R.~F., {Strom}, A.~L., {Steidel}, C.~C., \& {Rudie}, G.~C. 2016,
  \apj, 832, 171

\bibitem[{{Triani} {et~al.}(2020){Triani}, {Sinha}, {Croton}, {Pacifici}, \&
  {Dwek}}]{Triani20}
{Triani}, D.~P., {Sinha}, M., {Croton}, D.~J., {Pacifici}, C., \& {Dwek}, E.
  2020, \mnras, 493, 2490

\bibitem[{{van Dishoeck}(1990)}]{vDishoeck90}
{van Dishoeck}, E.~F. 1990, in Astronomical Society of the Pacific Conference
  Series, Vol.~12, The Evolution of the Interstellar Medium, ed. L.~{Blitz},
  207--228

\bibitem[{{van Dishoeck}(1992)}]{vDishoeck92}
{van Dishoeck}, E.~F. 1992, in Astrochemistry of Cosmic Phenomena, ed. P.~D.
  {Singh}, Vol. 150, 143

\bibitem[{{Van Loo} {et~al.}(2013){Van Loo}, {Butler}, \& {Tan}}]{VanLoo13}
{Van Loo}, S., {Butler}, M.~J., \& {Tan}, J.~C. 2013, \apj, 764, 36

\bibitem[{{Wolfire} {et~al.}(2022){Wolfire}, {Vallini}, \&
  {Chevance}}]{Wolfire22}
{Wolfire}, M.~G., {Vallini}, L., \& {Chevance}, M. 2022, \araa, 60, 247

\bibitem[{{Yasuda} {et~al.}(2023){Yasuda}, {Kuno}, {Sorai}, {Muraoka},
  {Miyamoto}, {Kaneko}, {Yajima}, {Tanaka}, {Morokuma-Matsui}, {Takeuchi}, \&
  {Kobayashi}}]{Yasuda23}
{Yasuda}, A., {Kuno}, N., {Sorai}, K., {et~al.} 2023, \pasj, 75, 743

\bibitem[{{Zanella} {et~al.}(2018){Zanella}, {Daddi}, {Magdis}, {Diaz Santos},
  {Cormier}, {Liu}, {Cibinel}, {Gobat}, {Dickinson}, {Sargent}, {Popping},
  {Madden}, {Bethermin}, {Hughes}, {Valentino}, {Rujopakarn}, {Pannella},
  {Bournaud}, {Walter}, {Wang}, {Elbaz}, \& {Coogan}}]{Zanella18}
{Zanella}, A., {Daddi}, E., {Magdis}, G., {et~al.} 2018, \mnras, 481, 1976

\bibitem[{{Zhao} {et~al.}(2024){Zhao}, {Liu}, {Zhang}, \& {Bisbas}}]{Zhao24}
{Zhao}, Y., {Liu}, J., {Zhang}, Z.-Y., \& {Bisbas}, T.~G. 2024, \apj, 977, 46

\bibitem[{{Zhukovska}(2014)}]{Zhukovska14}
{Zhukovska}, S. 2014, \aap, 562, A76

\end{thebibliography}

\appendix

\section{Density-weighted gas temperatures}
\label{app:Tgas}

\begin{figure}
    \centering
    \includegraphics[width=0.95\linewidth]{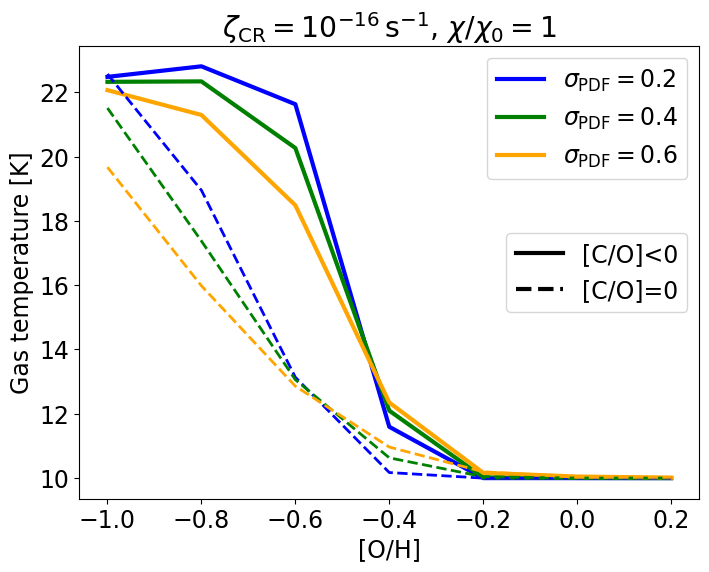}
    \caption{Density-weighted gas temperatures for the $A_{\rm V}$-PDF distributions with $m=15$ and $\sigma_{\rm PDF}=0.2$ (blue), 0.4 (green) and 0.6 (orange). In all cases, $\zeta_{\rm CR}=10^{-16}\,{\rm s}^{-1}$ and $\chi/\chi_0=1$. Solid lines represent the $\rm [C/O]<0$ case, whereas dashed lines the $\rm [C/O]=0$ case.}
    \label{fig:Tgas}
\end{figure}

Figure~\ref{fig:Tgas} shows the average, density-weighted, gas temperature of three $A_{\rm V}$-PDFs with $m=15$ and $\sigma_{\rm PDF}$ of 0.2 (blue line), 0.4 (green line) and 0.6 (orange line). In all cases $\zeta_{\rm CR}=10^{-16}\,{\rm s}^{-1}$ and $\chi/\chi_0=1$. The gas temperature increases with decreasing [O/H], which impacts the shape of CO SLEDs. For the aforementioned distribution, we find that $\langle T_{\rm gas}\rangle$ is $\sim10\,{\rm K}$ for $\rm [O/H]\sim-0.2$ to $0.2$, while for $\rm [O/H]\lesssim-0.8$ it can reach $\sim20-22\,{\rm K}$. This enhances the CO emission and  particularly the high-$J$ transitions which can eventually lead to elevated CO SLEDs.

With dashed lines we plot the linear case ($\rm [C/O]=0$ and $\cal D$ respectively). For metallicities $\rm [O/H]\gtrsim-0.4$, the average gas temperature does not show an appreciable difference. However, for lower metallicities the linear relation predicts a lower gas temperature, primarily due to the higher dust-to-gas ratios which attenuate the FUV radiation decreasing, consequently, the amount of photoelectric heating. However, for very metal-poor gas (e.g. $\rm [O/H]\sim-1.0$), the gas temperatures predicted by both bases of $\rm [C/O]$ are quite similar. 

\section{Linear assumption of [C/O] and $\cal D$ versus [O/H]}
\label{app:linear}

\begin{figure*}
    \centering
    \includegraphics[width=0.48\textwidth]{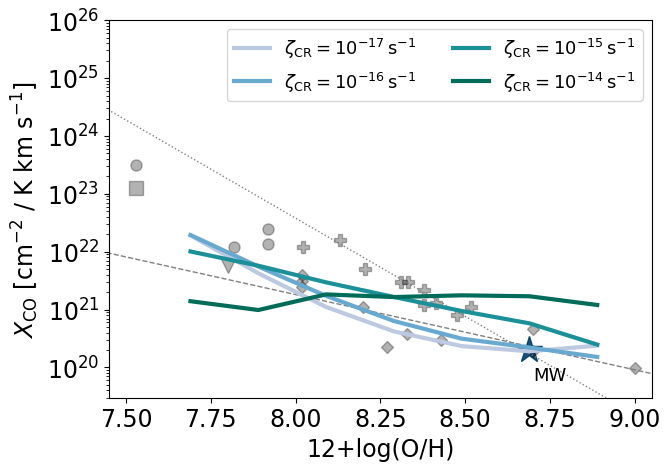}
    \includegraphics[width=0.48\textwidth]{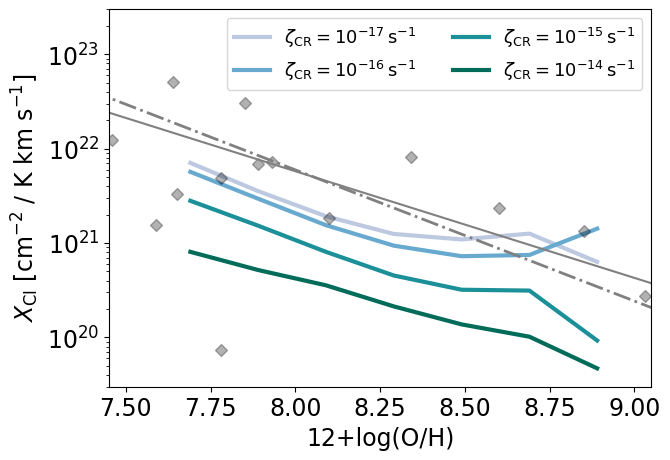}
    \includegraphics[width=0.48\textwidth]{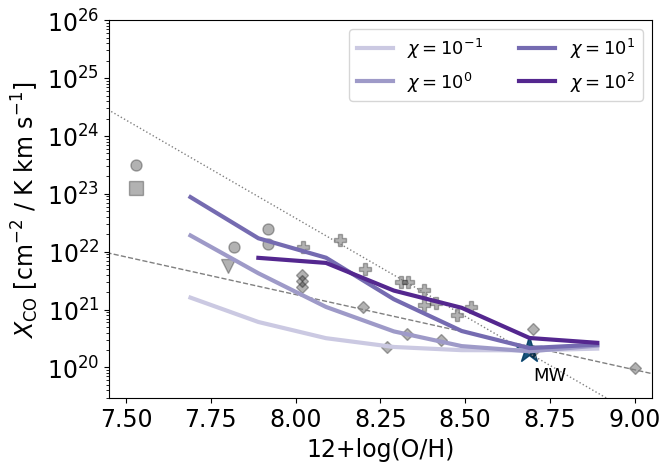}
    \includegraphics[width=0.48\textwidth]{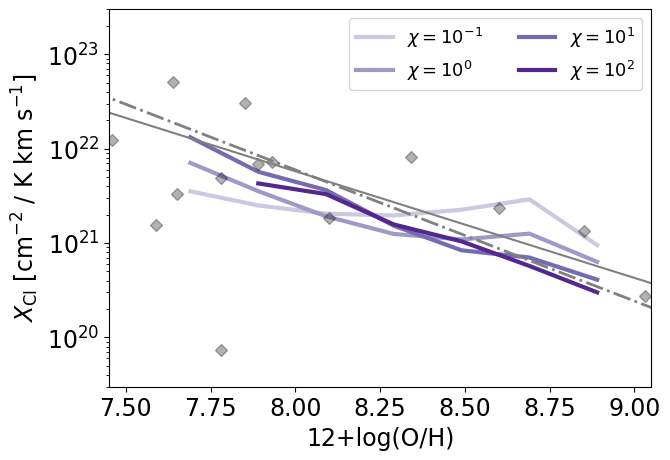}
    \includegraphics[width=0.48\textwidth]{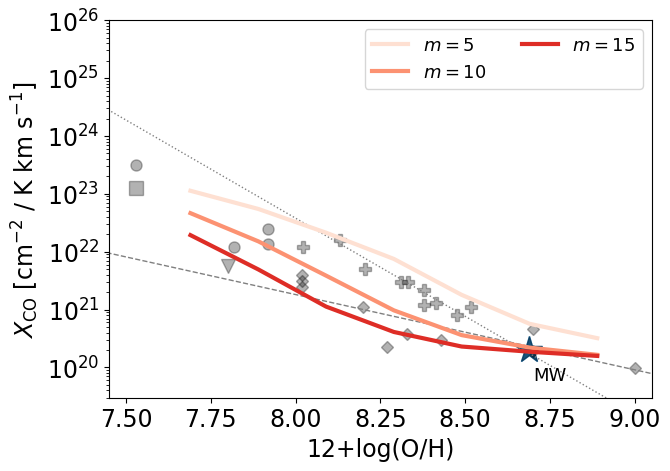}
    \includegraphics[width=0.48\textwidth]{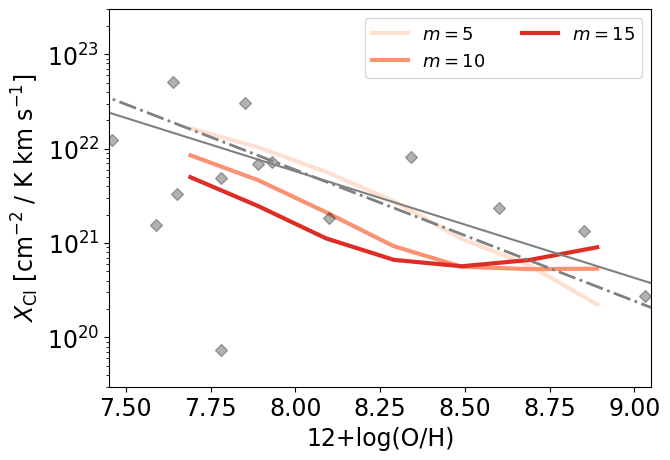}
    \caption{As in Figure~\ref{fig:xfacdep} for the ``linear" conditions of C/O ratio ($\rm [C/O]=0$) and the dust-to-gas ratio ($\cal D$).} 
    \label{fig:xfacdep-lin}
\end{figure*}

Table~\ref{tab:linear} shows the abundances of carbon, oxygen and the gas-to-dust ratio (normalized to the solar value) when a linear decrease versus metallicity is assumed. Figure~\ref{fig:xfacdep-lin} (similar to Fig.~\ref{fig:xfacdep}) shows the dependence of both $X$-factors on the cosmic-ray ionization rate, the FUV intensity and the $A_{\rm V}$-PDF versus metallicity for that assumption.

Under the linear assumption, the $X_{\rm CO}$-factor shows a less pronounced increase with decreasing metallicity compared to the non-linear case, particularly at low metallicities ($\rm [O/H]\lesssim -0.6$). This is because the linear decrease in [C/O] and $\cal D$ results in less extreme variations in dust shielding and CO formation efficiency. However, $X_{\rm CO}$ still exhibits significant sensitivity to $\zeta_{\rm CR}$ and $\chi/\chi_0$, with higher cosmic-ray ionization rates leading to non-linear fluctuations in $X_{\rm CO}$ due to changes in gas temperature and CO formation pathways.

For $X_{\rm CI}$, the linear assumption results in a more gradual decline with increasing metallicity compared to the non-linear case. The [C{\sc i}]-to-H$_2$ factor remains relatively stable across a wide range of metallicities, with less sensitivity to $\zeta_{\rm CR}$ and $\chi/\chi_0$ than $X_{\rm CO}$. This concludes that [C{\sc i}] (1-0) is a more reliable tracer of molecular gas in low-metallicity environments, even under simplified assumptions about the ISM composition.

The linear assumption, while useful for comparison, underestimates the complexity of the ISM environment and its impact on molecular gas tracers.

\begin{table}[]
    \centering
    \begin{tabular}{c|c|c|c}
        [O/H] & $\cal D$ & C & O  \\ \hline
        0.2 & 1.58 & $4.26(-4)$ & $7.76(-4)$ \\
        0      & 1.00 & $2.69(-4)$  & $4.90(-4)$ \\
        $-0.2$ & 0.63 & $1.69(-4)$ & $3.09(-4)$ \\
        $-0.4$ & 0.39 & $1.07(-4)$ & $1.95(-4)$ \\
        $-0.6$ & 0.25 & $6.76(-5)$ & $1.23(-4)$ \\
        $-0.8$ & 0.16 & $4.25(-5)$ & $7.77(-5)$ \\
        $-1.0$ & 0.10 & $2.69(-5)$ & $4.90(-5)$ \\
    \end{tabular}
    \caption{As in Table~\ref{tab:ics} but assuming a linear decrease of $\rm [C/O]$ and $\cal D$ (dotted black lines in both panels of Fig.~\ref{fig:ics}.}
    \label{tab:linear}
\end{table}

\end{document}